\documentclass[prb, twocolumn]{revtex4}

\usepackage{graphicx}

\usepackage{amsmath}
\usepackage{dsfont}
\usepackage{amstext}
\usepackage{amssymb}
\usepackage{amsfonts}
\usepackage{amsbsy}
\usepackage{bbm}
\usepackage{amsthm}
\usepackage{graphicx}
\usepackage{color}
\renewcommand{\v}[1]{\boldsymbol{#1}}
\renewcommand{\t}[1]{\tilde{#1}}

\newcommand{\e}{\hspace{1pt}\mathrm{e}}
\newcommand{\dd}{\hspace{1pt}\mathrm{d}}
\newcommand{\imth}{\hspace{1pt}\mathrm{i}\hspace{1pt}}

\newcommand{\Ref}[1]{Ref.~\onlinecite{#1}}

\newcommand{\eq}[1]{(\ref{#1})}
\newcommand{\eqn}[1]{eqn.~(\ref{#1})}

\newcommand{\<}{\langle}
\renewcommand{\>}{\rangle}

\newcommand{\prt}{\partial}

\newcommand{\ie}{{\it ie~}}

\newcommand{\etc}{{\it etc~}}

\newcommand{\al}{\alpha}

\newcommand{\del}{\delta}
\newcommand{\Del}{\Delta}
\newcommand{\eps}{\epsilon}

\newcommand{\ga}{\gamma}

\newcommand{\la}{\lambda}

\newcommand{\om}{\omega}

\renewcommand{\th}{\theta}

\newcommand{\si}{\sigma}

\newcommand{\cB}{ {\cal B} }

\newcommand{\cE}{ {\cal E} }

\newcommand{\cH}{ {\cal H} }

\newcommand{\cL}{ {\cal L} }

\newcommand{\cR}{ {\cal R} }

\newcommand{\bpm}{\begin{pmatrix}}
\newcommand{\epm}{\end{pmatrix}}
\newcommand{\bmm}{\begin{matrix}}
\newcommand{\emm}{\end{matrix}}

\newcommand{\kp}[1]{\noindent \emph{
 \begin{itemize}
 #1
 \end{itemize}
}
}

\begin{document}

\title{
Emergence of helicity $\pm2$ modes (gravitons) from qubit models }
\author{Zheng-Cheng Gu}
\author{Xiao-Gang Wen}
\affiliation{
Department of Physics, Massachusetts Institute of Technology,
Cambridge, Massachusetts 02139
}
\date{{\small April 2007}}
\begin{abstract}

Gauge symmetry is commonly regarded as one of the founding
principles of nature.  But recent studies of
topological/quantum order suggest that gauge symmetry can
emerge as a low energy property of a qubit model that have
no gauge symmetry at all at lattice scale. This suggests
that gauge symmetry may not be a founding principle, but
merely a property of a quantum ground state with long range
quantum entanglements.

The general equivalence principle and the associated diffeomorphism gauge
symmetry are also regarded as a founding principles of nature. So one may
wonder, can diffeomorphism gauge symmetry also emerge as a low energy property
of certain topological/quantum order in a qubit model?  In this paper, we
designed qubit models (or quantum spin models) on 3D lattice and showed that,
for the first time, gapless helicity $\pm 2$ excitations (\ie the gravitons)
can emerge as the \emph{only} low energy excitation in our models. We showed
that the emergence of gapless helicity $\pm 2$ excitations in our models leads
to the emergence of, at least, linearized diffeomorphism gauge symmetry
$h_{\mu\nu}\to h_{\mu\nu} +\prt_\mu f_\nu+\prt_\nu f_\mu$ at low energies.

In the first qubit model (called the L-type model), we show
that helicity $\pm 2$ gapless excitations appear as the
\emph{only} type of low energy excitations using a reliable
semiclassical approach.  The dispersion of the gapless
helicity $\pm 2$ is found to be $\eps_{\v k} \propto |\v
k|^3$. The appearance of the gapless helicity $\pm2$ modes
suggests that the ground state of the qubit model is a new
state of matter.  In the second model (called the N-type
model) the collective modes are strongly interacting and
there is no reliable approach to understand its low energy
dynamics.  Using a spin-wave/quantum-freeze approach (which
is shown to reproduce the correct emergent $U(1)$ gauge
theory in a quantum rotor model), we argue that the second
model may contain helicity $\pm 2$ gapless excitations as
the only type of low energy excitations with a linear
dispersion $\om \propto k$.

Both models have emergent low energy diffeomorphism gauge
symmetry which leads to the associated gapless helicity $\pm
2$ excitations. We believe that those properties are
topologically robust: any translation invariant
perturbations cannot break the emergent diffeomorphism gauge
symmetry and cannot generate a gap for those helicity $\pm
2$ excitations.  Our results shed light on the quest to find
a quantum theory of gravity -- a quantum model with a finite
cutoff whose ground state supports gravitons (\ie helicity
$\pm 2$ gapless excitations with a linear $\eps_{\v k}
\propto |\v k|$ dispersion).

\end{abstract}

\maketitle

{\small \setcounter{tocdepth}{1} \tableofcontents }

\section{Emergence approach}

\subsection{Discoveries and unifications}

\kp{\item
The development of theoretical physics is driven by
the desire to understand everything from a single or very few origins.
}

Although the down pull by the earth was realized even before human
civilization, such a phenomena did not arose any curiosity.  After
Galileo and Kepler found that planets move in a certain particular
way described by a mathematical formula, people started to wonder:
why planets move in such a peculiar and precise way.  This motivated
Newton to develop his theory of gravity.  Newton's theory not only
explains the planets motion, it also explains the down-pull that we
feel on earth.  It unifies the two seemingly unrelated phenomena.
Later we discovered that two other seemingly unrelated phenomena,
electricity and magnetism, can generate each other.  Our curiosity
about the electricity and magnetism leads to another giant leap in
science, which is summarized by Maxwell equations. Maxwell theory
unifies electricity and magnetism and reveals that light is merely
an electromagnetic wave. We gain a much deeper understanding of
light, which is so familiar and yet so unexpectedly rich and complex
in its internal structure.  Newton's theory of gravity and Maxwell's
theory of light illustrate how science develops: it develops through
the cycles of discovery, unification, more discovery, more
unification.

\subsection{Seven wonders of universe}

\kp{\item
The current physical understanding of our world is build on
seven basic assumptions.
\item Field theory and geometry played the central role
in formulating the fundamental theory of nature.
\item
This paper is a non-geometric approach to gravity
(and other wonders).
}

Modern science has made many more discoveries and has also unified
many seemingly
unrelated discoveries into a few simple structures.  Those simple structures
are so beautiful and we regard them as wonders of our universe.  They are also
very mysteries since we do not understand where do they come from and why do
they have to be the way they are.  At moment, the most fundamental mysteries
and/or wonders in our universe can be summarized by the following short list:\\
(1) Locality.\\
(2) Identical particles.\\
(3) Gauge interactions.\cite{Wey52,P4103,YM5491}\\
(4) Fermi statistics.\cite{F2602,D2661}\\
(5) Chiral fermions.\cite{LY5654,Wo5713}\\
(6) Lorentz invariance.\cite{E0591}\\
(7) Gravity.\cite{E1669}

In the current physical theory of nature, we take the above
properties for granted and do not ask where do they come
from.  We put those wonderful properties into our theory by
hand, for example, by introducing one field for each kind of
interactions or elementary particles.

Here, we would like to question where those wonderful and
mysterious properties come from.  Following the trend of
science history, we wish to have a single unified
understanding of all of the above mysteries.  Or more
precisely, we wish that we can start from a single structure
to obtain all of the above wonderful properties.

Historically, the first attempt to unify different forces is
motivated by Einstein general relativity which view gravity
as an distortion of space.\cite{E1669} In 1918, Weyl
proposed that the unit that we used to measure physical
quantities is relative and is defined only locally.  A
distortion of the unit system can be described by a vector
field which is called gauge field.  Weyl proposed that such
a vector field (the gauge field) is the vector potential
that describes the electromagnetism.  Although the above
particular proposal turns out to be incorrect, the Weyl's
idea is correct.  In 1925, the complex quantum amplitude was
discovered.  If we assume the complex phase is relative,
then a distortion of unit system that measure local complex
phase can also be described by a vector field. Such a vector
field is indeed the vector potential that describes the
electromagnetism.  This leads to a unified way to understand
gravity and electromagnetism: gravity arises from the
relativity of spacial directions at different spatial
points, while electromagnetism arises from the relativity of
complex quantum phases at different spatial points.
Nordstr\"om, M\"oglichkeit, Kaluza, and Klein further showed
that both gravity and electromagnetism can be understood as
a distortion of space provided that we think the space as
five dimensional with one dimension compactified into a
small circle.\cite{NM1404,K2166,K2695} Since that time, the
geometric understanding of gravity and electromagnetism have
dominated theoretical physics.

In this paper,\cite{GW0600} we will take a position that the geometric
understanding is not good enough and will try to advocate a very
different non-geometric understanding of gravity and electromagnetism.
Why the geometric understanding is not good enough? First the
geometric understanding is not self-consistent.  The consideration
based quantum mechanics and Einstein gravity indicates that two points
separated by a distance less than the Planck length cannot exist as a
physical reality. Thus the foundation of the geometric approach --
manifold -- simply does not exist in our universe. This suggests that
geometry is an emergent phenomenon that appears only at long
distances. So we cannot use geometry and manifold as a foundation to
understand fundamental physical problems.

Maxwell theory of light and Einstein theory of gravity are built on top of
geometry. They fail to answer what is the origin geometry.  In other words,
Maxwell theory and Einstein theory predict light waves and gravitational
waves. But the theories fail to tell us what is waving?
Also, Maxwell theory and Einstein theory fail to answer what is the origin
of the other wonders of our universe, such as Fermi statistics.

\subsection{Locality principle and unification}

\kp{\item
Quantum theory and locality suggest a very natural
fundamental build block -- qubit.
But, can qubits unify everything?
}

In this paper we try to address the above question by trying to find a single
microscopic origin for both light and gravity.  We hope the microscopic origin
will allow us to gain a deeper and a unified understanding of gravity and
light. We also hope that a deeper and unified understanding of gravity and
light will lead to a unified understanding of other deep mysteries of the
universe, such as the origin of all elementary particles.

At first sight, it appears that such a goal is too ambitious
to be practical. However, recent progresses suggest that
such an ambitious goal may be achievable.  It was shown
that, starting from a single origin -- a local bosonic model
(which is also called a quantum spin model, or more
precisely a qubit model), the first four of the seven
wonders can emerge naturally at low energies, if the ground
state of the spin model is described by string-net
condensation.\cite{LWuni,Wqoem,LWqed} Thus we may say that
the string-net condensation in a qubit model
provides a unified origin for the four wonders: identical
particles, gauge interactions, Fermi statistics and near
masslessness of the fermions.  Two seemingly unrelated
properties, Gauge interaction and Fermi statistics, are
unified under the string-net picture.  It is amazing to see
that using emergence and the string-net picture to
understand the origin of light leads to a understanding of
the origin of Fermi statistics.

Knowing the above result, the goal of this paper can be
stated more precisely: in addition to the four emergent
properties mentioned above, we want to understand merely one
more emergent property -- the emergence of gravity, from a
local qubit model.  In this paper, we want to construct a
local qubit model where gravitational waves and gravitons
emerge as the low energy excitations of the model.  If
successful, such a local qubit model will actually represent
a quantum theory of gravity.  So constructing such a local
qubit model solves a long standing problem of putting
quantum mechanics and gravity together.  We hope in the
future, we can construct a local qubit model where light,
gravity, and nearly massless fermions emerge at low
energies.  Such a local qubit model will unify five of the
seven deep mysteries in our list.

The belief that all the wonderful phenomena of our universe (such as
gauge interaction, Fermi statistics, gravitons, or even superstring
theory) emerge from a lowly qubit model is
called locality principle.\cite{Wqoem} Using a local qubit model as
a underlying structure to understand the deep mysteries of our
universe\cite{Wen04,LWuni} represents a departure from the traditional
approach to understand our world by dividing things into smaller parts
(which will be called division approach). It also  represents a
departure from the geometric way to understand gravity and gauge
interactions.

\subsection{Two approaches}

\kp{\item
In the traditional division approach, people are looking for
the fundamental build block of matter, by dividing particles into smaller and smaller pieces.
\item
In the emergence approach, we are looking for
the fundamental build block of space itself and assume that
such building block is qubit. The elementary particles
are viewed as excitations of qubits that from the space.
}

In the division approach, we try to understand various things by divide them
into smaller and smaller parts.  If we assume the division has to end at a
certain level, then we conclude that all things are formed by the parts that
cannot be divided further.  The indivisible parts are call the elementary
particles.  So in the division approach, we view everything in our world as
made of some simple beautiful building blocks, the elementary particles.  A
deeper understanding is gained if we find some elementary particles are not
actually elementary and are formed by even smaller objects.  A large part of
science is devoted in finding those smaller and smaller objects, as
represented by the discoveries of atoms, electrons and protons, and then
quarks.

However, the division approach that we followed in last 150 years may
not represent a right direction.  For example, phonons in a solid is
as particle-like as any other elementary particles at low energies.
But if we look at phonons closely, we do not see smaller parts that
form a phonon.  We see the atoms that fill the entire space.  The
phonons are not formed by those atoms, the phonons are simply
collective motions of those atoms.

This leads us to wonder that maybe photons, electrons, gravitons, \etc, are
also collective motions of a certain underlying structure that fill the entire
space. They may not have smaller parts. Looking for the smaller parts of
photons, electrons, and gravitons to gain a deeper understanding of those
elementary particles may not be a right approach.

In this paper, we will use a different approach, emergence approach, to gain a
deeper understanding of elementary particles.  In the emergence approach used
in this paper, we view space as a collection of qubits. The empty space (the
vacuum) corresponds to the ground state of the qubits, and the elementary
particles (the matter formed by them) correspond to the excitations of the
qubits.

We know that elementary particles in our would can have quite different
properties.  Can excitations of qubits have those rich properties?  Due to the
particle-wave duality in quantum theory, particles and waves are the same
thing. In the emergence approach we will try to understand the nature of waves
in order to understand the nature of elementary particles.

\subsection{Originate from organization}

\kp{\item
In the emergence approach,
the particular properties of elementary particles
do not come from the particular choice of building blocks
(there is only one choice in the emergence approach -- the qubit),
but come from how qubits are organized in the ground state.}

To see how the emergence approach leads to a deeper understanding of elementary
particles, let us start with Euler equation $\prt_t^2\rho-v^2\prt_x^2\rho=0$
that describe waves in a liquid. We want to ask what is the microscopic origin
of Euler equation?  The answer is obtained only after the discovery of atoms.
We find that the waves in a liquid are collective motions of the atoms and
Euler equation describes the dynamics of the those collective motions. However,
how atoms are organized in the ground state plays an important role here. Only
when the atoms are organized into a liquid state (or more precisely, a boson
condensed state for the bosonic atoms), are the collective motions of those
bosonic atoms described by Euler equation. If the bosons organize into a
crystal, then their collective motions will be described by a different wave
equation, Navier equation: $\prt_t^2 u^i-T^{ijk}_l \prt_j\prt_k u^l=0$.

Due to the particle-wave duality, our understanding of Euler equation and
Navier equation leads to an understanding of phonons in liquids and solids.
We see that it is the \emph{organization} of bosons that becomes the
microscopic origin of phonons. Different organizations lead to different wave
equations and different types of phonons.

{}From this point of view, to find a microscopic origin of light and
gravity is to find a particular organization of bosons (or
qubits),\footnote{A lattice bosonic model and a lattice qubit model (\ie
a lattice spin model) are equivalent.  So in this paper the two terms,
bosons and qubits, are equivalent and are used interchangeably.} such
that the collective motions of such organized bosons (qubits) are
described by Maxwell equation and Einstein equation (instead of Euler
equation or Navier equation).  We like to point out that, in such an
emergence approach, we do not look for the building blocks of
everything. The building block is known: it is a qubit.  Even the
detail form of the qubit model and the precise values of various
coupling constants are not important. The important issue is how the
bosons (qubits) are organized in the ground state.  Different
organizations lead to different collective excitations, or in other
words, different sets of ``elementary'' particles.

It was shown that
if bosons (qubits) organize into network of strings and if the
string-nets form a quantum liquid (called string-net condensed state),
then photons, electrons and quarks can emerge naturally as collective
motions of such organized bosons
(qubits).\cite{LWuni,Wqoem,LWqed} In this paper, we will try to
find an organization of bosons (qubits) such that the collective
motions of bosons (qubits) also lead to gravitons.

In contrast, in the traditional approach, we choose a particular field for each
kind of elementary particle.  The particular properties of elementary particles
are encoded in our particular choice of the field.  For example, to get a
fermion (such as electron), we choose an anti-commuting Grassmann field, while
to get a photon, we choose a vector field.

We like to point out that this paper uses a particular emergence
approach: we try to obtain everything from a local qubit model
Another emergence approach was developed in
superstring theory\cite{GreSW88} in last 10 years, as demonstrated by
the duality relations among various superstring models and matrix
models.\cite{Pol98} The anti-de Sitter-space/conformal-field-theory
duality even shows how space-time and gravity emerge from a gauge
theory.\cite{AGM0083} It would be interesting find out if the
gravity from the superstring theory can produce helicity $\pm2$ modes
(gravitons) \emph{as the only gapless excitation}.  It is also
interesting to find a relation between all those ideas of emergence.

\section{The strategy of our approach}

\subsection{The rule of game}

\kp{\item
A practical definition of quantum gravity.
}

There are many different approaches to quantum
gravity\cite{T0294,C0185,P9487} based on different principles. Some
approaches, such as loop quantum gravity,\cite{S0448} stress the
gauge structure from the diffeomorphism of the space-time and try to
build a quantum theory where the whole space-time is emergent. (This
is called background independence.\cite{S0535}) Other approaches,
such as superstring theory,\cite{GreSW88,Pol98} stress the
renormalizability of the theory and try to construct a
renormalizable theory that contain gravitons.  In this paper, we
will stress different things, namely finiteness and locality, and
follow a different rule of game.

Our rule of game is encoded in the following working definition of quantum
gravity.  Quantum gravity is\\
(a) A quantum theory described by a Hamiltonian in a Hilbert space.\\
(b) Its Hilbert space has a finite dimension.\\
(c) The Hamiltonian is a sum of local operators.\\
(d) A single gapless helicity $\pm 2$ mode is the only type of
low energy excitations.\\
(e) The  helicity $\pm 2$ excitations have a linear dispersion.\\
(f) The gravitons \footnote{In this paper, we define gravitons as linearly
dispersing gapless helicity $\pm 2$ excitations.  }
interact in the way consistent with experimental observations.\\
We like to remark that different people may have different
understandings/definitions of quantum gravity. We hope the above definition
that will be used in this paper will help to avoid some possible confusions.

Is the above definition of  quantum gravity a proper definition? The
first problem that we face when we consider quantum gravity is that
``is quantum gravity a quantum theory?''\footnote{Here, quantum theory
is defined as a theory whose states are state vectors in a Hilbert
space, and observables and Hamiltonian are Hermitian linear operators.
So in quantum theory, by definition, there exists an absolute time.
[We note that path integral formulation can some times go beyond
quantum mechanics (say when time is discretized).]} The answer to the
above question appears to be NO, since Einstein theory tells us that
time is dynamical while quantum mechanics assume an absolute time.

The second problem is that ``is quantum gravity a theory based real
and complex numbers?'' As mention above, the consideration of quantum
theory and Einstein gravity suggests that the notion of manifold is
not a physical reality. Similarly, one may wonder, maybe even real
numbers and complex numbers cannot exist as a physical reality. In
this case, we wonder: should we avoid using real and complex numbers
when we formulate a theory of quantum gravity?

The above two considerations may serve as a long term goal. But they
are not so useful as a guide for research. This is why we proposed a
\emph{working} definition of quantum gravity to guide our research at this
stage. When we understand quantum gravity better, we expect that our
working definition will be modified.

Let us discuss each item in our working definition in more detail.  Since the
time is absolute in quantum mechanics, the condition (a) implies that the time
is not emergent.  So the quantum gravity considered here is not background
independent.  In order for the time to be emergent, one has to go beyond
quantum mechanics, that we will not do in this paper.

The condition (b) implies that the quantum gravity considered here has a
finite cut-off.  So the renormalizability is not an issue.  A condensed matter
system always has a finite cut-off and can only simulates a system with a
finite cut-off. So the conditions (a--b) makes it possible to use condensed
matter system to simulate the quantum gravity (as defined here).  Also, from a
mathematical point of view, only theories with finite cut-off are really well
defined.

The condition (c) is a locality condition.  It implies two additional
things: (1) the total Hilbert space is a direct product of local
Hilbert spaces $\mathcal{H}_{tot} = \otimes_i \mathcal{H}_i$.  (2) the
local operators are defined as operators that act within each local
Hilbert space $\mathcal{H}_i$ or finite products of local operators.
The conditions (a--c) actually define a local bosonic model (or a
local qubit model).\cite{Wqoem} Certainly, any quantum spin models
satisfy (a--c).  It is the condition (d--f) that makes a theory to
look like gravity.

But why do we require gapless gravitons to be the only low energy excitations in
the condition (d)?  To answer this question, let us consider the theory of
electromagnetism where gapless photons (excitations with helicity $\pm 1$) are
the only low energy excitations.  If a system also has helicity $0$ gapless
excitations, then we are no longer sure if the gapless helicity $\pm 1$
excitations are photons.  This is because phonons in solid have helicity $0$
and $\pm 1$.  Thus a theory with gapless helicity $0$ and helicity $\pm 1$
excitations may not be a theory of photons, it may be a theory of phonons.
But a theory with  helicity $\pm 1$ excitations as the only low energy
excitations must be a theory of electromagnetism.  So here we require our
theory to have helicity $\pm 2$ excitations as the only low energy
excitations, to make sure the theory is a theory of quantum gravity (at least
at linearized level).

The condition (d) is a very important condition.
It is very easy to construct a quantum model that contain helicity
$\pm 2$ gapless excitations, such as the theory described by the following
Lagrangian for symmetric tensor field
\begin{equation*}
 \cL=
\frac12 \prt_t h^{ij}\prt_t h^{ij}
-\frac12 \prt_k h^{ij}\prt_k h^{ij} .
\end{equation*}
Such a theory also contain helicity $0$ and $\pm 1$ gapless excitations and is
certainly not a theory of gravity.  So here we impose the condition (d) rule
out the above example.  It is highly non-trivial to construct a quantum model
that contains helicity $\pm 2$ excitations as the only gapless excitations.
Many theories of quantum gravity fail this test.

We would like to mention that, according to our definition,
only when helicity $\pm 2$ modes are the only low lying
excitations, can the theory be a theory of quantum gravity.
Such a condition may be too strict. We may want to relax the
condition (d) to condition (d'): the helicity $\pm 1$ and
$\pm 2$ excitations are the only low energy excitations. In
this case the helicity $\pm 1$ excitations are photons and
the helicity $\pm 2$ excitations are gravitons, which
reflects the situation in our universe.

The emergent gravitons from local qubit models  naturally
interact with each other.  However, they in general interact
in a different way from that described by the higher order
non-linear terms in Einstein gravity.  Since those higher
order terms in Einstein gravity are irrelevant at low
energies and not universal when viewed from the perspective
of local qubit models,  therefore it may be possible to
generate those higher order terms by fine tuning the lattice
model [such as modifying the Hamiltonian ($H_J$ and $H_g$),
the constraints ($H_U$), as well as the Berry's phase term
in \eqn{Lthphi}].  So it may be possible that local qubit
models can generate proper non-linear terms to satisfy (f).

To summarize, the goal of our approach to quantum gravity is
to construct an Hamiltonian operator that act on a Hilbert
space.  We require that the Hilbert space is a directly
product of local Hilbert spaces which each has a finite
dimension: $\cH=\otimes_i \cH_i$.  Respect to such a
locality structure of the Hilbert space, we require that the
Hamiltonian operator is a sum of local operators.  (The
local operators are defined as operators that act within
each local Hilbert space $\mathcal{H}_i$ or finite products
of local operators.) In order for the Hamiltonian operator
to describe a quantum theory of gravity, we also require
that all the low energy excitations above the ground state
of the Hamiltonian are described by a single mode of
helicity $\pm 2$ excitations (gravitons).  Last, those
gravitons should interact in way that is required by the
equivalence principle of Einstein gravity.

\subsection{A brief outline of our approach}

\kp{\item
How to design a qubit model whose only gapless excitations are
described by a single helicity $\pm 2$ mode.
}

We first start with a field theory of a symmetric tensor
$h_{ij}$ described by the following phase space Lagrangian:
\begin{align*}
 \cL&=\pi^{ij}\prt_0 h_{ij}
-J_1\pi^{ij} \pi^{ij}
-J_2\pi^{ii} \pi^{jj}
\nonumber\\
&\ \ \
-g_1\prt_k h_{ij} \prt_k h_{ij}
-g_2\prt_i h_{ij} \prt_k h_{kj}
-g_3\prt_i h_{ij} \prt_j h_{kk}.
\end{align*}
We then put the theory on the lattice to have a finite
cut-off.  The key issue is that, as a quantum theory, does
the lattice model has helicity $\pm 2$ modes as the only
gapless excitation?

It turns out that it is very hard to have helicity $\pm 2$
modes as the only gapless excitation. In general, one either
has all helicity $\pm 2$, $\pm 1$, and $0$ modes as gapless
excitations or have no gapless excitation at all.  In this
paper, we show that if we compactify and descretize $h_{ij}$
and its canonical conjugate $\pi^{ij}$, we can actually have
a lattice model that has helicity $\pm 2$ modes as the only
gapless excitation.  Such helicity $\pm 2$ modes correspond
to the emergent gravitons from the lattice model.  We also
find that the low energy effective theory for such a lattice
model has an emergent linearized diffeomorphism gauge
symmetry $h_{\mu\nu}\to h_{\mu\nu} +\prt_\mu f_\nu+\prt_\nu
f_\mu$, which is an implied consequence of having helicity
$\pm 2$ modes as the only gapless excitation.

\subsection{Some previous approaches}

\kp{\item
There are many previous approaches to quantum gravity.  But only a few
of them produce a local quantum Hamiltonian (which satisfies
the conditions (a - c)).  Among those that produce a local
quantum Hamiltonian, non of them were shown to contain a
single helicity $\pm 2$ mode as the only gapless
excitation.}

In this section, we will discuss some previous approaches in
terms of our practical definition of quantum gravity.
Superstring theory\cite{GreSW88,Pol98} satisfies the
conditions (a), (e) and (f), but in general not (d) due to
the presence of dilatons (massless scaler particles).  The
superstring theory (or more precisely, the superstring field
theory) also does not satisfy the condition (b) since the
cut-off is not explicitly implemented.

Many approaches to quantum gravity are based on quantizing the classical Einstein
action.  The lattice gravity approach based Regge calculus on 4D space-time
lattice\cite{R6158,MT8161} belong to this type of approach where a finite
space-time cut-off is introduced.  Such an approach may not satisfy the
conditions (a-c) since it may not produce a local Hamiltonian and local Hilbert
space.  Such an approach eventually failed due to the doubling phenomenon of
the gravitational modes, \ie it does not satisfies the condition (d).

In \Ref{RS8975,WZ9489,DM0779}, quantum gravity is studied in terms of lattice
Hamiltonians with a continuous time (just like this paper).  However, those
Hamiltonians obtained by quantizing the classical Einstein action are non-local
(\ie they do not satisfy the condition (b,c)).  Also, the low energy excitations
of the those lattice Hamiltonians were not discussed.  We do not know if those
lattice Hamiltonians have gapless excitation or not.  As a result, we do not
know if those lattice Hamiltonians give rise to emergent diffeomorphism gauge
symmetry at low energies.  The spin network\cite{RS9543} or the quantum
computing\cite{S0535} approach to quantum gravity satisfies the condition (a,b)
or (a--c).  But again, the properties (d--f) remain to be shown.

The induced gravity from superfluid $^3$He discussed in \Ref{V9867} does not
satisfy the condition (d) due to the presence of gapless superfluid mode.  In
\Ref{ZH0123}, it is proposed that gravitons may emerge as edge excitations of a
quantum Hall state in 4 spatial dimensions. Again the condition (d) is not
satisfied due to the presence of helicity $\pm S$ modes where $S$ is unbounded.
In fact, there are infinite many gapless modes with various helicities.

In  \Ref{X0643}, a very interesting spin model is constructed.  The
model satisfies the condition (a -- c) and\\
(d'') the gapless
helicity $0$ and helicity $\pm 2$ modes are the only low energy
excitations;\\
(e'') the helicity $\pm 2$ modes have a quadratic
dispersion.\\
The model is interesting since its ground state is a new
state of matter -- an algebraic spin liquid. Such a state is beyond
the Landau's symmetry breaking description. Despite the emergence of
gapless helicity $\pm 2$ modes, due to the properties (d'') and (e''),
the low energy properties of the model are not very close to those of
Einstein gravity.

In this paper, we will try to fix the two problems and try to
construct a qubit model that satisfies the condition (a--d),
and hopeful (f) through fine tuning.  We only partially achieve our
goal.

We studied two quantum spin models (or qubit models).  In the first
quantum spin model (called the L-type model), the helicity $\pm 2$
gapless excitations are reliably shown to appear as the only type of low
energy excitations (\ie the conditions (a--d) are satisfied). Within
a perturbative calculation, the dispersion of the gapless helicity
$\pm 2$ is found to be $\eps_{\v k} \propto |\v k|^3$.  So the
condition (e) is not satisfied.  The appearance of the gapless
helicity $\pm2$ modes suggests that the ground state of the quantum
spin model is a new state of matter.

In the second model (called the N-type model)\cite{GW0600} the
collective modes are strongly interacting and there is no reliable
approach to understand its low energy dynamics.  Using a
spin-wave/quantum-freeze approach (which is shown to reproduce the
correct emergent $U(1)$ gauge theory in a quantum rotor model), we
argue that the second model may contain helicity $\pm 2$ gapless
excitations as the only type of low energy excitations with a linear
dispersion $\om \propto k$.  More reliably numerical calculations are
needed to confirm that the type-N model really has $\om \propto k$
helicity $\pm 2$ excitations as the only low energy excitations.

We believe that the gaplessness of the helicity $\pm 2$ excitations in
both models is topologically robust: any translation invariant
perturbations cannot generate a gap for those  helicity $\pm 2$
excitations.

\section{Review of U(1) gauge theory}

\label{sec:u1}

In this paper, we are going to construct a qubit model with
emergent helicity $\pm2$ gapless modes from a theory of
symmetric tensor.  We will show that by imposing some
``constraints'' through certain spin interaction terms, we
can obtain a theory where the helicity $\pm 2$ excitations
are the only low energy excitations.  Since the constraints
that we will impose are similar to the Gauss constraint in
$U(1)$ gauge theory, here we will first give a brief review
of quantum $U(1)$ gauge theory, to introduce the physical
ideas behind our construction in a more familiar setting.

To obtain a $U(1)$ gauge theory, we may start with continuum quantum
field theory of vector fields $a_i$ and $\mathcal{E}^i$. The phase
space Lagrangian has a form
\begin{equation}
\label{Lvec}
 \mathcal{L}=
 - \mathcal{E}^i \partial_0a_i
- \frac12 J \mathcal{E}^i \mathcal{E}^i
- \frac12 g_1 \partial_i a_j \partial_i a_j
- \frac12 g_2 \partial_i a_i \partial_j a_j
\end{equation}
{}From the equation of motion
\begin{equation}
 \partial_0^2 a_i=
Jg_1 \partial^2 a_i
+Jg_2 \partial_i \partial_j a_j
\end{equation}
we find the low energy excitations to be helicity $\pm 1$ and
helicity $0$ modes with linear dispersions $\omega \propto k$.
Despite the presence of gapless helicity $\pm 1$ modes, the above
theory is not a theory of electromagnetism due to the presence of
gapless helicity $0$ mode. So the key to obtain a theory of
electromagnetism is to gap the helicity $0$ mode.

\subsection{Removing helicity 0 mode though Gauss constraint}

\kp{\item
In the standard field theory approach, we obtain photons by
starting with a vector field theory, and then removing the
helicity $0$ mode by imposing a $U(1)$ gauge symmetry and
the resulting Gauss constraint.
\item
The Gauss constraint makes the Hilbert space non-local
(\ie violate the condition (c)).}

The standard way to remove the helicity $0$ mode at low
energies  is to impose Gauss constraint
\begin{equation}
 \partial_i \mathcal{E}^i=0\label{M1}
\end{equation}
In the constraint system, $a_i$ becomes a many-to-one label of the
physical states.
Different vector fields related by an local transformation
\begin{equation}
 a_i\rightarrow a_i+\partial_i f\label{M2}
\end{equation}
actually label the same state.  This is the well known gauge transformation
which is generated by the Gauss constraint.  Physical quantities should be
invariant under such transformation so that the same state always has the same
values of physical quantities.

In particular, the Lagrangian for $a_i$ field should be gauge
invariant. An easy way to do this is to write down the Lagrangian in
terms of gauge invariant fields. The gauge invariant fields are the
magnetic field
\begin{equation}
\mathcal{B}^i=\epsilon^{ijk}\partial_j a_k\label{M3}
\end{equation}
and the electric field $\cE^i$.
The Lagrangian is then
\begin{equation}
 \mathcal{L}=
 - \mathcal{E}^i \partial_0a_i
- \mathcal{H}(a_i,\mathcal{E}^i)
\label{M4}
\end{equation}
\begin{equation}
 \mathcal{H}=
\frac{1}{2}[J (\mathcal{E}^i)^2+ g (\mathcal{B}^i)^2]
\label{M5}
\end{equation}

The resulting equation of motion contains only two transverse modes
corresponding to the helicity $\pm 1$ excitations. The Gauss
constraint removes the helicity $0$ excitations from the low energy
spectrum.

However, removing helicity $0$ excitations through the Gauss
constraint has one problem. The Hilbert space of the physical states
is modified by the constraint, since only $\mathcal{E}^i$ that
satisfies the Gauss constraint are physical and only gauge
inequivalent $a_i$ correspond to different physical states.  The new
Hilbert space can no longer be written as a direct product of local
Hilbert spaces, since an arbitrary local change of  $\mathcal{E}^i$ in
general violate the constraint.  Thus the resulting quantum system is
no longer a local qubit system.

\subsection{Try to gap helicity 0 mode though energy penalty}

\kp{\item
Using an energy penalty
to impose the Gauss constraint fails, since it
fails to gap the helicity 0 mode.}

To fix this problem, here we choose not to impose the constraint and not to
change the Hilbert space. We choose instead to include a term of form $U
(\partial_i \mathcal{E}^i)^2$ in the Hamiltonian. The resulting
theory is described by
\begin{equation}
 \mathcal{L}=
 - \mathcal{E}^i \partial_0a_i
- \mathcal{H}(a_i,\mathcal{E}^i)
-\frac12 U(\partial_i \mathcal{E}^i)^2
\label{M4a}
\end{equation}
We hope that new term will suppress the fluctuations that violate
the Gauss constraint and will gap the helicity $0$ excitations.
However, from the equation of motion obtained from \eq{M4a}
\begin{equation}
 \partial_0^2 a_i=
g (J-U\prt^2) ( \partial^2 \del_{ij}  -\prt_i\prt_j) a_j ,
\end{equation}
we find that the extra $\frac12 U(\partial_i \mathcal{E}^i)^2$ term
cannot gap the helicity $0$ mode.  The helicity $0$ mode remains to
have zero velocity.

\subsection{Another failed attempt}

\kp{\item
Imposing Gauss constraint through an energy penalty
fails to gap helicity 0 mode even on lattice.}

Next, we will put the continuum theory \eq{M4a} on lattice and examine
if the $U(\partial_i \mathcal{E}^i)^2$ term can gap the helicity $0$
mode on lattice.  To put the theory \eq{M4a} on a cubic lattice, we
introduce $a_{\v i\v j}$ and $\cE_{\v i\v j}$ for each link $\<\v i\v j\>$
of the cubic lattice.  Here $\v i$ labels the sites of the cubic
lattice and $a_{\v i\v j}$ and $\cE_{\v i\v j}$ satisfy $a_{\v i\v
j}=-a_{\v j\v i}$ and $\cE_{\v i\v j}=-\cE^{\v j\v i}$. The phase
space Lagrangian for physical degrees of freedom $a_{\v i\v j}$ and
$\cE_{\v i\v j}$ is given by
\begin{align}
\label{lattL}
&\mathcal{L} =
 \sum_{\<\v i\v j\>} \cE_{\v i\v j} \prt_0 a_{\v i\v j}
-\frac J2 \sum_{\<\v i\v j\>} \cE_{\v i\v j}^2
-\frac g2 \sum_{\<\v i\v j\v k\v l\>} \cB_{\v i\v j\v k\v l}^2
-\frac U2 \sum_{\v i} Q_{\v i}^2
\nonumber\\
& \cB_{\v i\v j\v k\v l} = a_{\v i\v j}+ a_{\v j\v k}+ a_{\v k\v l}+
a_{\v l\v i},
\nonumber\\
& Q_{\v i} = \sum_{\v j\text{ next to }\v i} \cE_{\v i\v j} ,
\end{align}
where $\sum_{\v i}$ sum over all sites, $\sum_{\v i\v j}$ sum over
all links, and $\sum_{\v i\v j\v k\v l}$ sum over all square faces
of the cubic lattice. This  phase space Lagrangian tells us that
$a_{\v i\v j}$ and $\cE_{\v i\v j}$ form a canonical
momentum-coordinate pair.  The $U\sum_{\v i} Q_{\v i}^2$ term is
the lattice version of the $U(\partial_i \mathcal{E}^i)^2$ term.
The $g \cB_{\v i\v j\v k\v l}^2$ term and the $J\sum_{\<\v i\v
j\>} \cE_{\v i\v j}^2$ term correspond to the $(\cB^i)^2$ term and
the $(\cE^i)^2$ term in the continuum theory.  The lattice
equation of motion can be easily derived from the Lagrangian
\eq{lattL}. We find that there are still three gapless modes with
helicity $0$ and $\pm 1$ in the long-wave-length limit. The
$U\sum_{\v i} Q_{\v i}^2$ term cannot gap the helicity $0$ mode
even on lattice.

\subsection{Gapping helicity 0 modes on lattice with discretized $\cE_{\v i\v
j}$ (or campactified $a_{\v i\v j}$)}
\label{clssU1}

\kp{\item
Imposing Gauss constraint through an energy penalty can gap
helicity 0 mode, if we (i) put the theory on lattice, and
(ii) compactify the vector field $a_{i}$.}

All those failed attempts reveal the difficulty of gaping the helicity
0 mode without gaping helicity $\pm 1$ modes nor changing the Hilbert
space. In the following, we will show that in order to gap the
helicity $0$ mode within the same local Hilbert space, we must (A) put
the theory \eq{M4a} on lattice (which is done in \eqn{lattL}) and (B)
discretize $\cE_{\v i\v j}$ (or compactify $a_{\v i\v j}$).

How to discretize $\cE_{\v i\v j}$?  From \eq{lattL} we see that for a
fixed link $\v i\v j$, if we view $a_{\v i\v j}$ as the coordinate
of a particle on a line then $\cE_{\v i\v j}$ is the momentum of the
particle.  To discretize (or quantize) the momentum $\cE_{\v i\v
j}$, we simply put the particle on a circle instead of a line. This
is achieved by letting $a_{\v i\v j}$ and $a_{\v i\v j}+2\pi$ to
describe the same point (thus to compactify $a_{\v i\v j}$).  After
the compactification, $\cE_{\v i\v j}$ is quantized as integer and
the phase space Lagrangian \eq{lattL} need to be modified to
\begin{align}
\label{lattLC}
L &=
 \sum_{\<\v i\v j\>} \cE_{\v i\v j} \prt_0 a_{\v i\v j}
-\frac12 J\sum_{\<\v i\v j\>} \cE_{\v i\v j}^2
+g\sum_{\<\v i\v j\v k\v l\>} \cos(\cB_{\v i\v j\v k\v l})
\nonumber\\
&\ \ \
-\frac12 U\sum_{\v i} Q_{\v i}^2
\end{align}
in order to be consistent with the periodic condition $a_{\v i\v j}\sim a_{\v
i\v j}+2\pi$.  \eq{lattLC} describes a rotor model with rotors on the links of
a cubic lattice. It was shown that in the $U\gg g \gg J$ limit, the ground
state of the rotor model is a string-net condensed
state.\cite{Walight,Wen04,LWqed} The low energy excitations above the
string-net condensed state are shown to be gapless helicity $\pm 1$ modes and
the helicity $0$ mode is gapped!\cite{FNN8035,MS0204,Walight,Wen04} Basically, after quantization, the operator
$\sum_{\v i} Q_{\v i}^2$ has a discrete spectrum. So in the large $U$ limit,
the term $U\sum_{\v i} Q_{\v i}^2$ remove some fluctuations from the low
energy spectrum. The removed fluctuations turn out to be the helicity $0$
mode.

\subsection{Low energy collective modes of the rotor model
through spin-wave/quantum-freeze approach
}

\label{U1spinwv}

\kp{\item
When we treat our lattice model as a classical theory, we find some classical
gapless modes have very weak quantum fluctuations, and they remain gapless in
quantum theory.  While other classical gapless modes have very strong quantum
fluctuations, and they acquire a gap in quantum theory.  }

The gapping of the helicity $0$ mode can also be understood from a
spin-wave approach if we incorporated a mechanism called quantum
freeze.  We will use a similar approach to study the emergence of
helicity $\pm2$ gapless mode.

To understand the low energy dynamics of the complicated and
strongly interacting rotor model \eq{lattLC}, let us treat
the model as a classical model
and $(a_{\v i\v j}, \cE_{\v i\v
j})$ as classical fields.  The classical ground state is obtained
by minimizing the Hamiltonian
\begin{align*}
\frac12 J\sum_{\<\v i\v j\>} \cE_{\v i\v j}^2 -g\sum_{\<\v i\v j\v k\v
l\>} \cos(\cB_{\v i\v j\v k\v l})
+\frac12 U\sum_{\v i} Q_{\v i}^2.
\end{align*}
We find that the classical ground state is given by $(a_{\v i\v j},
\cE_{\v i\v j})=(0,0)$.  The classical low energy collective modes is
given by the fluctuations $a_{\v i\v j}$ and $\cE_{\v i\v j}$.

One way to obtain the dynamics of the
classical  collective modes is to obtain
the continuum
effective theory of lattice model \eq{lattLC}.  To
obtain the continuum theory, let us assume the
fluctuations of $a_{\v i\v j}$ are
small and expand \eq{lattLC} to the quadratic order of $a_{\v i\v j}$.  Then we
take the continuum limit by introducing two vector fields
$(\v a,\v\cE)$ and
identifying
\begin{align*}
 a_{\v i\v j} &=\int_{\v i}^{\v j}\dd  x^i a^i ,
&
 \cE_{\v i\v j} &=\int_{\v i}^{\v j}\dd  x^i \cE^i ,
\nonumber\\
 \cB_{\v i\v j\v k\v l} &=\int_{\v i\v j\v k\v l}\dd  \v S\cdot
(\v \prt\times \v a)
\end{align*}
where $\int_{\v i\v j\v k\v l}\dd  \v S$ is the surface integration
on the square $\v i\v j\v k\v l$, and we have assumed that the
lattice constant $a=1$.  The resulting continuum effective theory is
given by \eq{M4}.  The classical collective modes described by $\v
a$ and $\v \cE$ are the spin-wave-like fluctuations in the rotor
model.  We find that there are three gapless modes with
helicity 0 and $\pm 1$ at classical level.

Now the question is that do we trust the above classical spin wave
result? So in the following, we will study the quantum
fluctuations of those classical modes to check the self
consistency of the classical spin-wave approach.

To study the quantum fluctuations of $\cE^i$ and $a_i$, we note that
the longitudinal mode and the transverse modes separate.  Introduce
$\v \cE=\cE_{||}+\cE_{\perp}$ and $\v a=a_{||}+a_{\perp}$, we find
that the dynamics of the transverse mode is described by
\begin{equation*}
 \cL_{\perp}=
a_{\perp}\prt_0 \cE_{\perp}
-\frac J2 \cE_{\perp}^2 - \frac g2 \prt_i a_{\perp} \prt_i a_{\perp}
\end{equation*}
At the lattice scale $\del x\sim 1$, the quantum fluctuations
of $\cE_{\perp}$ and $a_{\perp}$ are given by
\begin{equation*}
 \delta\cE_{\perp}\sim \Big(\frac gJ\Big)^{1/4},\ \ \ \ \ \
 \delta a_{\perp}\sim \Big(\frac Jg\Big)^{1/4}.
\end{equation*}
We see that when $J\ll g$ the fluctuations of $\v a$ is much less
than 1.  So expanding $\cos(B_{\v i\v j\v k\v l})$ to quadratic
order is a good approximation. But this alone does not grantee the
validity of the spin-wave approach.  In quantum theory, the
compactness of $a_{\v i\v j}$ imply that $\cE_{\v i\v j}$ is
discrete. So we cannot treat $\cE_{\v i\v j}$ as a continuous
variable, as we did in the classical spin-wave approach.  However,
in the $J\ll g$ limit, we see that the quantum fluctuations of the
$\v \cE$ (or $\cE_{\v i\v j}$) is much larger than 1 which is the
discreteness of $\cE_{\v i\v j}$. In this case, we can indeed
treat $\cE_{\v i\v j}$ as continuous variables.  So the spin-wave
approach is valid in the  $J\ll g$ limit for the transverse mode.
The classical spin wave result can be trusted even in quantum
theory.  We conclude that the transverse mode (or the helicity
$\pm 1$) has a linear gapless dispersion.

The longitudinal mode is described by two scalar fields
$(f(\v x), \pi(\v x))$
with $a_i=\prt_i f$ and $\pi
=\prt_i\cE^i$. Its dynamics is determined by
\begin{equation*}
\cL_{||}= \pi \prt_0 f
-\frac J2 \pi(-\prt^{-2})\pi -\frac 12 U \pi^2.
\end{equation*}
At the lattice
scale, the quantum fluctuations of $\pi$ and $f$ are given by $\del
\pi = 0$ and $\del f = \infty$.  We see that
the fluctuations of $f$ are much bigger than the compactification size $2\pi$
and the fluctuations of $\pi$ are much less then the discreteness of $\cE^i$
which is $1$.  In this limit, the result from the classical spin-wave approach
cannot be trusted once we include quantum effect.

In fact the weak quantum fluctuations in the discrete
variable $\pi$ and the strong quantum fluctuations in the compact
variable $f$ suggest that the corresponding mode is gapped after
the quantization. Such a phenomenon is called quantum freeze.

Since $\pi$ has weak fluctuations which is less
than the discreteness of $\pi$, the ground state is basically given
by $\pi=0$.
A low lying
excitation is then given by $\pi=0$ everywhere except in a unit cell
where $\pi = 1$.  Such an excitation have an energy of order $U$. The
gapping of helicity $0$ mode is confirmed by more
careful calculations.\cite{Wen04,LWqed} From those calculations, we find
that the weak fluctuations of $\pi$ lead to a constraint $
\pi=\prt_i\cE^i=0$ and the strong fluctuations of $f$ lead to a
gauge transformation $ a_i\to a_i+\prt_i f$ for the low energy states.
The Lagrangian \eq{M4a}
equipped with the above constraint and the gauge transformation
becomes the Lagrangian of a $U(1)$ gauge theory.

To summarize, the emergence of photons (the helicity $\pm 1$
excitations) from a local qubit model is
purely a quantum effect which requires a discretization of the
electric field on lattice. It also requires the ground state of the
qubit model to have a new kind of order -- string-net
condensation (which is ensured by the condition $J\ll g$).  In this
case, the helicity $\pm 1$ modes have small fluctuations and are
gapless, while the helicity $0$ mode has extremely strong quantum
fluctuations and are gapped. This example gives us important hints on
what are required in order to have emergent helicity $\pm 2$
excitations from a qubit model.

\section{Generalization to symmetric tensor field}
\label{tensor}

\subsection{A  field theory of symmetric tensor}

\kp{\item
The phase-space Lagrangian of a symmetric-tensor
field theory.}

To construct a qubit model with emergent helicity $\pm 2$
gapless modes, let us
start with a continuum field theory with symmetric tensor fields
$a_{ij}$ and $\cE^{ij}$
\begin{equation*}
 a_{ij}=a_{ji},\ \ \ \  \ \ \
\cE^{ij}=\cE^{ji} ,
\end{equation*}
where $i,j=1,2,3$ are the spatial indices.
 The phase space Lagrangian of the field theory
is given by
\begin{align*}
 \cL&=\cE^{ij}\prt_0 a_{ij}
-J_1\cE^{ij} \cE^{ij}
-J_2\cE^{ii} \cE^{jj}
\nonumber\\
&\ \ \
-g_1\prt_k a_{ij} \prt_k a_{ij}
-g_2\prt_i a_{ij} \prt_k a_{kj}
-g_3\prt_i a_{ij} \prt_j a_{kk}.
\end{align*}
We see that $\mathcal{E}^{ij}$ is the canonical momentums of $a_{ij}$
and satisfy
\begin{align}
\label{comm}
 [a_{ij}(\v y),\cE^{mn}(\v x)]=\frac{i}{2}(\delta_{im}\delta_{jn}+\delta_{jm}\delta_{in})\del(\v x-\v y)
\end{align}
as operators.

The model can be put on a lattice and becomes a qubit model.
From the resulting equation of motion, we find that
there are six gapless linear modes.  Two modes with helicity $0$ and
four modes with helicities $\pm 1$ and $\pm 2$. So to obtain emergent
graviton, we simply need to gap the two helicity $0$ modes and the two
helicity $\pm 1$ modes while keep the helicity $\pm 2$ modes gapless.

\subsection{$\partial_i \cE^{ij}=0$ constraint
and helicity 0 and $\pm 2$ quadratic modes}

\kp{\item
The vector constraint and the associated gauge symmetry
remove the helicity $0$ and $\pm 1$ modes
in symmetric tensor field theory.}

First, let us try to remove those modes through
constraints in the continuum theory.
We first impose three constraints
\begin{equation}
\label{T1}
 \partial_i \mathcal{E}^{ij}=0
\end{equation}
which are called the vector constraints.
Obviously these three constraints commute
with each other.  From the constraint, we can construct a unitary operator
\begin{equation}
W = \e^{- 2i \int d^3\v x f_j(\v x) \prt_i\cE^{ij}(\v x)}
\label{TG}
\end{equation}
that generates gauge transformations on $a_{ij}$:
\begin{equation}
\label{T2}
 a_{ij}\rightarrow Wa_{ij}W^\dag =
a_{ij}+\partial_i f_j+\partial_j f_i
\end{equation}
The gauge invariant field is a symmetric tensor field
\begin{eqnarray}
\label{T3}
R^{ij}=R^{ji}=\epsilon^{imk}\epsilon^{jln}\partial_m
\partial_l a_{nk}
\end{eqnarray}

The dynamics of the constraint system is describe by a
gauge invariant Lagrangian.  To the lowest order, such Lagrangian has a form
\begin{eqnarray}
\cL=\cE^{ij}\prt_0a_{ij}-
{\al (\mathcal{E}^{ij})}^2-\beta{(\mathcal{E}^{ii})}^2-\ga
{(R^{ij})}^2-\la {(R^{ii})}^2\label{T5}
\end{eqnarray}
with $\alpha,\beta,\gamma,\la>0$. After solving the equations motion, one finds
that the constraint system contains three gapless modes with quadratic
dispersion $\om\propto k^2$. Two modes have helicities $\pm 2$ and one has
helicity $0$.  Thus the vector constraints \eq{T1} remove the two helicity $\pm
1$ modes and one helicity 0 mode.

In fact, this constraint model defined through \eq{T1} and \eq{T5} is
the continuum limit of the lattice model studied in \Ref{X0643}.  Due
to the quadratic dispersion and the helicity $0$ gapless mode, the
lattice model studied in \Ref{X0643} does not reproduce Einstein
gravity at low energies. However, the lattice model does give rise to
a new quantum liquid of qubits (or quantum spins).  Just like any other
ordered state, the  new quantum liquid also contain topological
defects.
As in U(1) gauge theory, we can allow the three constraints
(\ref{T1}) to be violated at one point
\begin{equation}
 \partial_i \mathcal{E}^{ij}(\v x)=q^j \del(\v x -\v x_0)\label{TB}
\end{equation}
and create a defect.
Such a defect is called ``electric'' charge.
It is clear that
the ``electric'' charge $q^i$ is a vector.
We also note that there are three identities
for $R^{ij}$
\begin{eqnarray}
\partial_i R^{ij}=0\label{T6}
\end{eqnarray}
in the continuum. Those identities can be violated on lattice. The
violation of these identities will also create a local defect called
the monopole. Again the corresponding ``magnetic'' charge is a
vector.

In Einstein gravity, the mass of matter is coupled to gravity via a
modification of constraint. In other words, the way that mass
generates gravity is similar to the way that electric or magnetic
charges generate electromagnetism. From the above discussion we see
that the model defined through \eq{T1} and \eq{T5} has no scaler
charges (electric or magnetic) that may correspond to mass.  This is
a more fundamental reason why the model can not be a theory of
quantum gravity.

\subsection{Additional $R^{ii}=0$ constraint
to remove the second helicity 0 mode}

\kp{\item
The additional scaler constraint and the associated gauge
symmetry remove the last helicity $0$ mode in symmetric
tensor field theory.
\item
In the constrained model, the gapless helicity $\pm 2$ modes
have a $\om \sim k^3$ dispersion.
\item
The scaler constraint can be violated at an isolated point,
which corresponds to a mass.
}

To obtain a theory with emergent gravitons and point scalar mass, we
would like to remove the second helicity $0$ mode by introducing another
constraint.  This constraint must be a scalar constraint.  There are
only two scalar physical constraints $\mathcal{E}^{ii}=0$ and $R^{ii}=0$ that
can be implemented in the theory.

If we use the constraint
\begin{eqnarray}
\mathcal{E}^{ii}=0\label{T8}
\end{eqnarray}
then the $R^{ij}$ is no longer gauge invariant under the gauge
transformation generated by the new constraint.  The lowest order
gauge invariant tensor field are third order derivatives of
$a_{ij}$.  The constraint does remove the helicity $0$ modes.  The
helicity $\pm 2$ modes have a cubic dispersion relation $\om \propto
k^3$.  Furthermore, we find that the ``electric charge'' cannot even
be point-like.  Only electric line charge is allowed in this case.
We see that the resulting theory is very different from Einstein
gravity.  So we will not go further in this direction, although the
resulting qubit system is a very interesting condensed
matter system which gives rise to a new quantum state of matter.

The right constraint that we should use is
\begin{eqnarray}
R^{ii}=0
\label{T9}
\end{eqnarray}
In terms of $a_{ij}$, this constraint is
\begin{eqnarray}
(\delta_{ij}\partial^2-\partial_i\partial_j) a_{ij}=0 \label{T10}
\end{eqnarray}
Violation of this constraint at one point corresponds to a scalar charge which
can be interpreted as a point mass.

But this constraint is very unusual, it is a
local constraint on $a_{ij}$ itself, so it generates a gauge
transformation on $\mathcal{E}^{ij}$
\begin{eqnarray}
\label{T11}
\mathcal{E}^{ij}\rightarrow \mathcal{E}^{ij}-(\delta_{ij}\partial^2
-\partial_i\partial_j)f_0
\end{eqnarray}
Under such a gauge transformation,
a linear gauge invariant field is given by
\begin{eqnarray}
\label{T12a}
C^i_j=\epsilon^{imn}\partial_m\left(\mathcal{E}^{nj}-\frac{1}{2}\delta_{nj}\mathcal{E}^{ll}\right)
\end{eqnarray}

Because $a_{ij}$ is the conjugate variable of $\mathcal{E}^{ij}$, we
should carefully check the commutating relation of these four
constraints. We first transform $a_{ij}, \mathcal{E}^{ij}$ into
Fourier space and it is easy to check
\begin{align}
\label{comm1}
 [a_{mn}(\v k),\cE^{ij}(\v k^\prime)]&=\frac{i}{2}[\del_{im}\del_{jn}+\del_{in}\del_{jm}]{(2\pi)}^3\del(\v k-\v k^\prime)
\end{align}
then we only need to check
\begin{eqnarray}
&&[k_j\mathcal{E}^{ij}(\v k), (\delta_{nm}{k^\prime}^2 -k_n^\prime
k_m^\prime)a_{nm}(\v k^\prime
)]\nonumber\\&&=\frac{i}{2}{(2\pi)}^3k_j(\delta_{nm}{k^\prime}^2
-k_n^\prime
k_m^\prime)[\del_{im}\del_{jn}+\del_{in}\del_{jm}]\del(\v k-\v
k^\prime)\nonumber\\
&&=i{(2\pi)}^3k_j(\delta_{ij}{k^\prime}^2 -k_i^\prime
k_j^\prime)\del(\v k-\v k^\prime)\nonumber\\
&&=0\label{T13}
\end{eqnarray}
It turns out that these four constraints commute with each other.


To obtain the dynamical properties of the symmetric tensor field
theory with the constraint \eq{T1} and \eq{T9}, we need to find a
gauge invariant Hamiltonian. One way to construct a gauge
invariant Hamiltonian is to express the Hamiltonian in terms of
gauge invariant fields $C^i_j$ in \eqn{T12a} and $R^{ij}$. This way
we obtain the following Hamiltonian density
\begin{eqnarray}
\label{T14}
\mathcal{H}=
\frac{J}{2} C^i_jC^i_j +\frac{g}{2} R^{ij}R^{ij}
\end{eqnarray}
or the Lagrangian density
\begin{align}
\label{LLtype}
 \cL&=\cE^{ij}\prt_0 a_{ij}
-\frac{J}{2} C^i_jC^i_j -\frac{g}{2} R^{ij}R^{ij} .
\end{align}
We will call such a system L-type model.

The equation of motion for this Hamiltonian is
\begin{eqnarray}
\dot{a}_{ij}&=&J\left(\epsilon^{imn}\partial_m C_j^n-\frac{1}{2}\delta_{ij}\epsilon^{nml}\partial_m C_l^n\right) \nonumber\\
\dot{\mathcal{E}}^{ij}&=&-g\epsilon^{nmj}\epsilon^{kli}\epsilon^{nm^\prime
k^\prime}\epsilon^{kl^\prime n^\prime}\partial_m
\partial_l\partial_{m^\prime}
\partial_{l^\prime} a_{{n^\prime}{k^\prime}}\label{T15}
\end{eqnarray}
If we use a plane wave solution along the third direction, with
$k_1=0, k_2=0, k_3=k$, it is easy to find
\begin{eqnarray}
\omega^2\mathcal{E}^{11}&=&gJ k^6\left(\mathcal{E}^{11}-\frac{1}{2}\mathcal{E}^{ll}\right)\nonumber\\
\omega^2\mathcal{E}^{22}&=&gJ k^6\left(\mathcal{E}^{22}-\frac{1}{2}\mathcal{E}^{ll}\right)\nonumber\\
\omega^2\mathcal{E}^{12}&=&gJ k^6\mathcal{E}^{12}\nonumber\\
\mathcal{E}^{13}&=&\mathcal{E}^{23}=\mathcal{E}^{33}=0 \label{T16}
\end{eqnarray}
We see that there are only two modes of gapless excitations,
$\mathcal{E}^{11}-\mathcal{E}^{22}$ and $\mathcal{E}^{12}$, with
$\om\propto k^3$ dispersions.  The fluctuations of form
$\cE^{11}+\cE^{22}$ has zero frequency and represent the pure gauge
fluctuations of \eq{T11}. So only the two helicity $\pm 2$ modes are
physical and the second helicity 0 mode is removed.

We can further calculate the force between two
scalar defects (the masses) obtained by violating the constraint
(\ref{T10}):
\begin{equation}
\label{masses}
 R^{ii} (\v x)=
m_1\del(\v x-\v x_1) +
m_2\del(\v x-\v x_2)
\end{equation}
We find the force is repulsive between two masses with the same sign
and decay as $|\v x_1-\v x_2|^{-4}$.

To summarize, we constructed a L-type model using symmetric fields
that satisfy the commutation relation \eq{comm} and the constraint
\eq{T1} and \eq{T9}. The gauge invariant Hamiltonian is given by
\eq{T17}. The only low energy excitations of the L-type model are
gapless helicity $\pm 2$ modes with a cubic dispersion.  Although
the  helicity $\pm 2$ excitations are the only low energy
excitations, the low energy properties of the L-type model are very
different from those of Einstein gravity where the  helicity $\pm 2$
excitations have a linear dispersion.

\subsection{Nonlocal gauge invariant Hamiltonian}

\kp{\item
In the constrained model, the gapless helicity $\pm 2$ modes can have a
linear $\om \sim k$ dispersion, if the field theory Lagrangian is gauge
invariant only up to a surface term. }

How to construct a model that has a linearly dispersing graviton and
$1/r^2$ attractive force between point masses with the same sign?
One way to achieve this is to choose a \emph{nonlocal} gauge
invariant Hamiltonian ({\it i.e.} the Hamiltonian density is gauge
invariant up to a total derivative). Let us consider the following
Hamiltonian density
\begin{eqnarray}
\label{T17}
\mathcal{H}=
\frac{J}{2}[{(\mathcal{E}^{ij})}^2 -\frac{1}{2}{(\mathcal{E}^{ii})}^2]\
+\frac{g}{2} a_{ij}R^{ij}
\end{eqnarray}
or the Lagrangian density
\begin{align}
\label{LNtype}
 \cL&=\cE^{ij}\prt_0 a_{ij}
-\frac{J}{2}[
\mathcal{E}^{ij} \mathcal{E}^{ij}
 -\frac{1}{2}{(\mathcal{E}^{ii})}^2]
-\frac{g}{2} a_{ij}R^{ij}
\end{align}
The quadratic term
\begin{eqnarray}
\mathcal{E}^{ij} \mathcal{E}^{ij}
-\frac{1}{2}{(\mathcal{E}^{ii})}^2
\label{T12}
\end{eqnarray}
is gauge invariant under (\ref{T11}), provided that $\cE^{ij}$ satisfy the
constraint \eq{T1} $\partial_{i}\mathcal{E}^{ij}=0$.  From the identity
(\ref{T6}), we also find that $a_{ij}R^{ij}$ is invariant under the gauge
transformation \eq{T2} up to a total derivative. Thus the total Hamiltonian is
invariant under the gauge transformation \eq{T11} and \eq{T2}.

The corresponding equations of motion have the form
\begin{eqnarray}
\omega^2\mathcal{E}^{11}&=&-gJ k^2\left(\mathcal{E}^{22}
-\frac{1}{2}\mathcal{E}^{ll}\right)\nonumber\\
\omega^2\mathcal{E}^{22}&=&-gJ k^2\left(\mathcal{E}^{11}
-\frac{1}{2}\mathcal{E}^{ll}\right)\nonumber\\
\omega^2\mathcal{E}^{12}&=&gJ k^2\mathcal{E}^{12}\nonumber\\
\mathcal{E}^{13}&=&\mathcal{E}^{23}=\mathcal{E}^{33}=0\label{T18}
\end{eqnarray}
{}From the solution of the above equations of motion, we find that
the only gapless modes are the two helicity $\pm 2$ modes with a
linear dispersion relation. Those modes can be identified as
gravitons after quantization.

We can create two point defects through
\eq{masses}.  We find that the force between the two masses is
proportional to $m_1m_2/r^2$.  Two masses with the same sign
attract and two masses with opposite signs repel.  The quantum
gravity theory defined through the commutation relation \eq{comm},
the constraint \eq{T1} and \eq{T9}, and non-local gauge invariant
Hamiltonian \eq{T17} is called N-type model. It is the N-type
model that represent a quantum theory of Einstein gravity (at
the linear level).

We would like to point out that when compared to Einstein theory
of gravity, $a_{ij}$ can be interpreted as the fluctuations of the
spatial part of a metric tensor $g_{\mu\nu}$ around flat space:
$a_{ij}\sim g_{ij}-\del_{ij}$. $R^{ij}$ is related to the three
dimensional Ricci tensor $\mathcal{R}^{ij}$:
\begin{eqnarray}
\mathcal{R}^{ij}=\frac{1}{2}(R^{ij}-\delta_{ij}R^{ll})\label{T4}
\end{eqnarray}

If we introduce $a_{00}$ and $a_{0i}$ as Lagrangian multipliers
to impose the vector and the scaler constraints,
we can rewrite our model (\ref{T14}) as the following
Lagrangian
\begin{eqnarray}
 \mathcal{L}&=&\mathcal{E}^{ij}\partial_0
a_{ij}-\frac{J}{2}\left[{(\mathcal{E}^{ij})}^2-\frac{1}{2}{(\mathcal{E}^{ii})}^2\right]-\frac{g}{2}
a_{ij}R^{ij} \nonumber\\ &&+ 2a_{0i}\partial_j
\mathcal{E}^{ij}+a_{00}(\partial^2 a_{ii}-\partial_i\partial_j
a_{ij})\label{T19}
\end{eqnarray}
After integrating out $\cE^{ij}$, we find that this action is
exactly the linearized Einstein action around a flat space-time.
The associate gauge transformations in space are also enlarged to
gauge transformations in space-time.

\section{The lattice models and imposing constraints through energy penalties}

\kp{\item
Put the symmetric-tensor field theory on lattice
and construct qubit models.}

So far, we have discussed continuum models that has gravitons as the only
gapless excitations.  The gapless helicity $0$ and $\pm 1$ modes are removed
by imposing the constraints \eq{T1} and \eq{T10}.  However, the two
constraints changes the Hilbert space of the model.  The new Hilbert space
does not have a local form $\cH=\otimes_n \cH_n$ where $\cH_n$'s are the local
Hilbert spaces.  Thus the constrained models are not local qubit models.
In this section, we are going to fix this problem. We will show that the
vector and the scalar constraints can be realized in lattice models through
certain energy penalty terms without changing the Hilbert space.

First let us put the symmetric tensor field theory on a cubic
lattice. \Ref{X0643} propose a nice way to do so. Here we will follow
that convention to put $a_{xx}$, $a_{yy}$, $a_{zz}$, $\cE^{xx}$,
$\cE^{yy}$, and $\cE^{zz}$ on the vertices and put $a_{xy}$,
$a_{yz}$, $a_{zx}$, $\cE^{xy}$, $\cE^{yz}$, and $\cE^{zx}$ on the
square faces of the cubic lattice. For example
\begin{align*}
 a_{xx}(\v x) & \to a_{xx}(\v i),
\nonumber\\
 a_{yy}(\v x) & \to a_{yy}(\v i),
\nonumber\\
 a_{zz}(\v x) & \to a_{zz}(\v i),
\nonumber\\
 a_{xy}(\v x) & \to a_{xy}(\v i+\frac{\v x}{2}+\frac{\v y}{2}),
\nonumber\\
 a_{yz}(\v x) & \to a_{yz}(\v i+\frac{\v y}{2}+\frac{\v z}{2}),
\nonumber\\
 a_{zx}(\v x) & \to a_{zx}(\v i+\frac{\v z}{2}+\frac{\v x}{2}),
\end{align*}
where $\v i$ is the integral vector that represents the position of a vertex of
the cubic lattice.

\subsection{Putting the constraints on lattice}

\kp{\item
The vector constraint and the scaler constraint are
put on lattice via an energy penalty term.
}

On the cubic lattice, one of vector constraints in \eqn{T1},
$\prt_i\cE^{ij}$, can be written in the following form
\begin{align*}
Q(\v i,\v i+\v x)=\ &  \cE^{xx}(\v i+\v x) - \cE^{xx}(\v i)
\nonumber\\
+&\cE^{yx}(\v i+\frac{\v x}{2}+\frac{\v y}{2})
-\cE^{yx}(\v i+\frac{\v x}{2}-\frac{\v y}{2})
\nonumber\\
+&\cE^{zx}(\v i+\frac{\v x}{2}+\frac{\v z}{2})
-\cE^{zx}(\v i+\frac{\v x}{2}-\frac{\v z}{2})
\end{align*}
The other two constraints become $Q(\v i,\v i+\v y)$ and $Q(\v i,\v
i+\v z)$ which are obtained from the above expression by cycling
$xyz$ to $yzx$ and $zxy$. We note that the constraint field
$\prt_i\cE^{ij}$ becomes a quantities on the links in the
$j$-direction. The scalar constraint \eq{T10} becomes
\begin{align*}
 \eta(\v i)= & \sum_{ a,b= x, y, z} [a_{bb}(\v i+\v  a) + a_{bb}(\v
i-\v  a) -2a_{bb}(\v i)]
\nonumber\\
- & \sum_{ a= x, y, z}
[a_{ a a}(\v i+\v  a) + a_{ a a}(\v i-\v  a) -2a_{ a a}(\v i)]
\nonumber\\
- & 2\times\sum_{ab=xy,yz,zx} [a_{ab}(\v i+\frac{\v a}{2}+\frac{\v
b}{2}) -a_{ab}(\v i-\frac{\v a}{2}+\frac{\v b}{2})
\nonumber\\
&\ \ \ \ \ \ \ \ \ \ \ -a_{ab}(\v i+\frac{\v a}{2}-\frac{\v b}{2})
+a_{ab}(\v i-\frac{\v a}{2}-\frac{\v b}{2})]
\end{align*}

Let us introduce, $L^{ij}$ and $\th_{ij}$. For example
\begin{align*}
 L^{xx}(\v i) &= (-)^{\v i} \cE^{xx}(\v i)
\nonumber\\
 L^{xy}(\v i+\frac{\v x}{2}+\frac{\v y}{2}) &= -(-)^{\v i} \cE^{xy}(\v i+\frac{\v x}{2}+\frac{\v y}{2})
\end{align*}
and
\begin{align*}
 \th_{xx}(\v i) &= (-)^{\v i} a_{xx}(\v i)
\nonumber\\
 \th_{xy}(\v i+\frac{\v x}{2}+\frac{\v y}{2}) &= -2(-)^{\v i} a_{xy}(\v i+\frac{\v x}{2}+\frac{\v y}{2})
\end{align*}
$L^{ab}$ and $\th_{ab}$ satisfy the following commutation relation
between angle and angular momentum:
\begin{align*}
 [L^{ab}(\v i),\th_{ab}(\v i)] &=-i, \ \ \ \
\text{others} = 0.
\end{align*}
Using $L^{ab}$ and $\th_{ab}$, we can rewrite the above discretized
constraints as
\begin{align*}
 Q(\v i,\v i+\v x) = &
 L^{xx}(\v i+\v x) + L^{xx}(\v i)
\nonumber\\
+&L^{yx}(\v i+\frac{\v x}{2}+\frac{\v y}{2})
+L^{yx}(\v i+\frac{\v x}{2}-\frac{\v y}{2})
\nonumber\\
+&L^{zx}(\v i+\frac{\v x}{2}+\frac{\v z}{2})
+L^{zx}(\v i+\frac{\v x}{2}-\frac{\v z}{2})
\end{align*}
and
\begin{align*}
 \eta(\v i) = &\ \sum_{ a,b= x, y, z}
[\th_{bb}(\v i+\v  a) +\th_{bb}(\v i-\v  a) +2\th_{bb}(\v i)]
\nonumber\\
- & \sum_{ a= x, y, z}
[\th_{ a a}(\v i+\v  a) +\th_{ a a}(\v i-\v  a) +2\th_{ a a}(\v i)]
\nonumber\\
- & \sum_{ab=xy,yz,zx}
[\th_{ab}(\v i+\frac{\v a}{2}+\frac{\v b}{2})
+\th_{ab}(\v i-\frac{\v a}{2}+\frac{\v b}{2})
\nonumber\\
&\ \ \ \ \ \ \ \ \ \ \
+\th_{ab}(\v i+\frac{\v a}{2}-\frac{\v b}{2})
+\th_{ab}(\v i-\frac{\v a}{2}-\frac{\v b}{2})]
\end{align*}
One can check that $Q(\v i,\v i+\v a)$ and $\eta(\v j)$ all commute
with each others.

Just as discussed in section \ref{sec:u1}, we would like to impose
the $\eta(\v i)=0$ and $Q(\v i,\v i+\v a)=0$ constraints by
including the term
\begin{equation*}
H_U'=  U_2\sum_{\v i} \eta^2(\v i)+
  U_1\sum_{\v i} \sum_{a=x,y,z} Q^2(\v i,\v i+\v a)
\end{equation*}
in the Hamiltonian without changing the Hilbert space.

However, using $H_U'$ to impose the $\eta(\v i)=0$ constraint does
not work. This is because $\eta$ has a continuous spectrum.  No
matter how large is $U_2$, the excitations that violate the $\eta(\v
i)=0$ constraint do not have an energy gap. So the low energy
excitations do not have to satisfy the $\eta(\v i)=0$ constraint.
Similarly, if $\th_{ab}$ are not compact and $L^{ab}$ are not
discrete, the $Q(\v i,\v i+\v a)$ operators will also have
continuous spectra.  The low energy excitations do not have to
satisfy the $Q(\v i,\v i+\v a)=0$ constraint.

\subsection{Compactification}

\kp{\item
To use the energy penalty to impose
the vector and scaler constraints, we need to
compactify and descretize the lattice fields
$ a_{xx}, a_{yy}, a_{zz}, a_{xy}, a_{yz}, a_{zx}$.
}

In order to use the energy penalty $H_U'$ to impose the $\eta(\v
i)=0$ and $Q(\v i,\v i+\v a)=0$ constraint, we need to make the
spectra of $\eta(\v i)$ and  $Q(\v i,\v i+\v a)$ discrete.  In this
case, for large $U_{1,2}$, all excitations below an energy gap of
order $U_{1,2}$ will satisfy the $\eta(\v i)=0$ and $Q(\v i,\v i+\v
a)=0$ constraint.  To discretize the spectrum of the constraint
operators $\eta(\v i)$ and $Q(\v i,\v i+\v a)$, we need to
discretize \emph{both} $L^{ab}$ and $\th_{ab}$ or compactify
\emph{both} $L^{ab}$ and $\th_{ab}$.

We compactify $\th_{ab}$ by imposing a periodic condition
$\th_{ab}\sim \th_{ab}+2\pi$.  Similarly, we compactify $L^{ab}$ by
imposing a periodic condition $L^{ab}\sim L^{ab}+n_G$ where $n_G$ is
an integer. $\th_{ab}$ and $L^{ab}$ are no longer physical operators
after the compactification.  Only $W_L^{ab}=\e^{2\pi i L^{ab}/n_G}$
and $W_\th^{ab}=\e^{i \th_{ab}}$ and their products are physical
operators.  For a fixed $ab$ and $\v i$, $W_L^{ab}(\v i)$ and
$W_\th^{ab}(\v i)$ satisfy the algebra
\begin{equation}
\label{thLalg}
 W_L^{ab}(\v i) W_\th^{ab}(\v i)
= \e^{2\pi i/n_G} W_\th^{ab}(\v i) W_L^{ab}(\v i)
\end{equation}
Such an algebra has only one $n_G$ dimensional representation.  This
$n_G$ dimensional representation becomes our local Hilbert space
$\cH_{\v i,ab}$. The total Hilbert space is given by
$\cH=\otimes_{\v i,ab} \cH_{\v i,ab}$ after the compactification. In
other words, there are $n_G^3$ states on each vertex and $n_G$
states on each square face of the cubic lattice.  The total
dimension of the Hilbert space is finite for a finite cubic lattice.
We note that in the $n_G\to \infty$ limit, we recover the
uncompactified case.

The constraint operators $\eta(\v i)$ and $Q(\v i,\v i+\v a)$ are
not allowed for the campactified model since they are not products
of $W_L^{ab}$ and $W_\th^{ab}$.  This can be fixed easily.  In the
campactified model we replace $\eta^2(\v i)$ and $Q^2(\v i,\v +\v
a)$ by $1-\cos[\eta(\v i)]$ and $n_G^2(1-\cos[2\pi Q(\v i,\v i+\v
a)/n_G])$ respectively.  So, in the compactified model, the terms in
the Hamiltonian that impose the constraints have the following form
\begin{align}
\label{HUC}
H_U &= n_G\t U_2\sum_{\v i} \{1-\cos[\eta(\v i)]\}
\\
 & +n_G\t U_1\sum_{\v i} \sum_{a=x,y,z} \{1-\cos[2\pi Q(\v i,\v i+\v a)/n_G]\}
\nonumber
\end{align}
The operators  $\cos[\eta(\v i)]$ and  $\cos[2\pi Q(\v i,\v j)/n_G]$ have
discrete eigenvalues.
We will also show below that those operators all commute with each other.
So the Hamiltonian $H_U$ has a finite energy gap.  The
low energy states below the gap ({\it i.e.} zero energy states) all satisfy
\begin{align}
\label{cmpcnstr} \e^{\imth \eta(\v i)}=\e^{2\imth \pi Q(\v i,\v i+\v
a)/n_G}=1.
\end{align}
We hope that $H_U$ in the
compactified model can gap all the helicity $0$ and $\pm 1$ modes.

\subsection{Low energy Hilbert space}
\label{lowspace}

\kp{\item
The structure of the low energy Hilbert space of the lattice model
that satisfies the vector and scaler constraints $H_U=0$.
}

\begin{figure}[tbp]
\begin{center}
\includegraphics[scale=0.5] {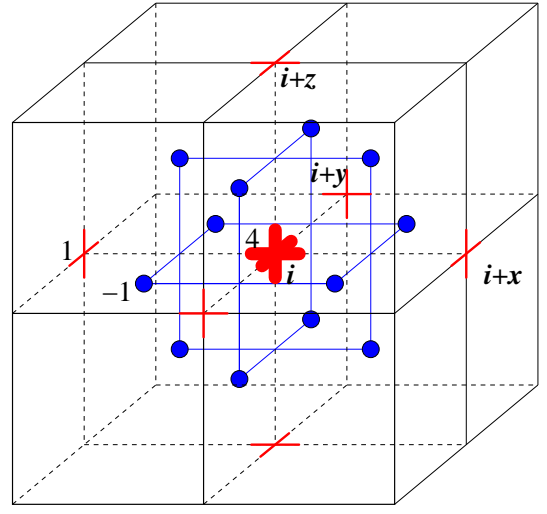}
\end{center}
\caption{
(Color online)
The integer coefficients $c^{ab}(\v i,\v r)$ in
\eq{Scabr}. A short solid line in the $a$ direction
 marks a non-zero  $c^{aa}(\v i,\v j)$.
 The thicker lines correspond to $c^{aa}(\v i,\v j)=4$ and
thinner lines $c^{aa}(\v i,\v j)=1$. The filled circles mark the
non-zero $c^{xy}(\v i,\v i+\frac{\v x}{2}+\frac{y}{2})$ \etc whose
values are equal to $-1$. Red represents positive integers and blue
negative integers. The action of the scaler constraint operator
$S(\v i)$ in \eqn{constr1} changes $L^{ab}(\v r)$ mod $n_G$. The
figure also shows those changes. The short solid lines mark the
non-zero changes of $L^{aa}(\v i)$ where thicker lines correspond
to a change of $4$ and thinner lines a change of 1. The filled
circles mark the non-zero changes of $L^{xy}(\v i+\frac{\v
x}{2}+\frac{y}{2})$ \etc. The changes are equal to $-1$. }
\label{eta}
\end{figure}

\begin{figure}[tbp]
\begin{center}
\includegraphics[scale=0.5] {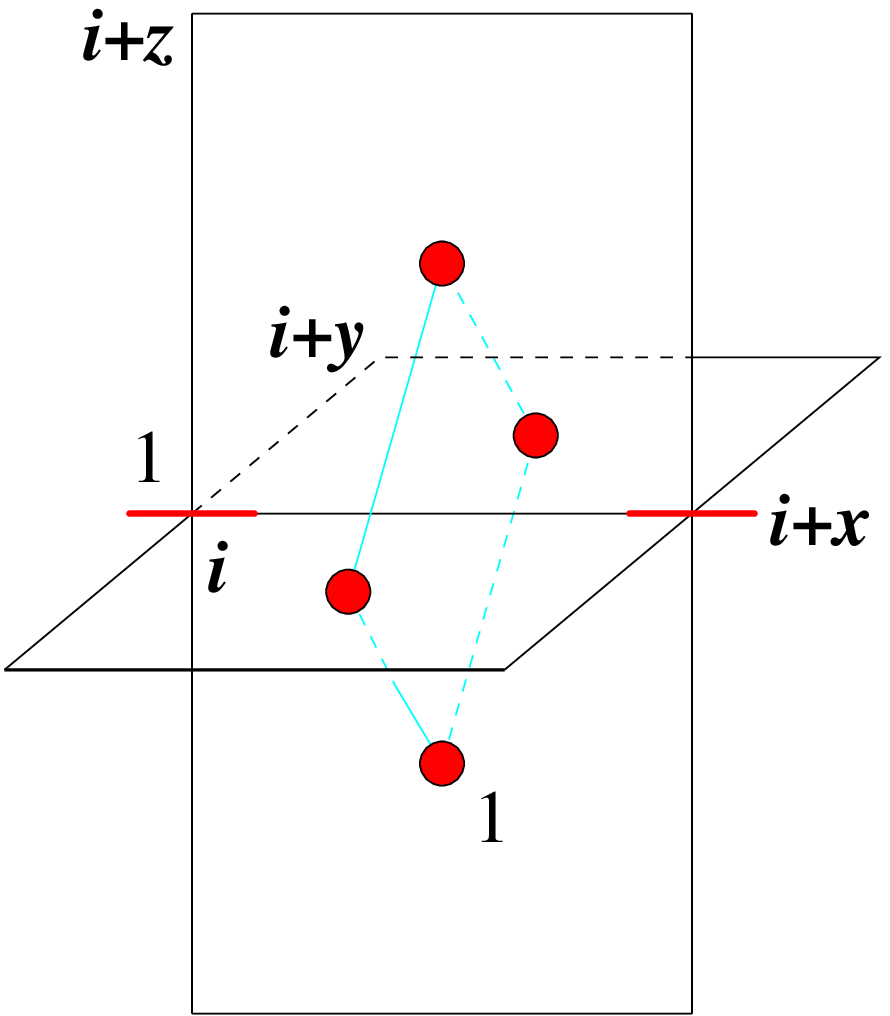}
\end{center}
\caption{
(Color online)
The integer coefficients $d^{ab}(\v i,\v i+\v x,\v r)$
in \eqn{Vdabr}. The short solid lines mark the non-zero  $d^{aa}(\v
i,\v i+\v x,\v j)$ which are equal to $1$. The filled circles mark
the non-zero $d^{xy}(\v i,\v i+\v x,\v i+\frac{\v
x}{2}+\frac{y}{2})$ \etc, which are also equal to $1$. The action
of the vector constraint operator $V(\v i,\v i+\v x)$ in
\eq{constr2} changes $n_G\th^{ab}(\v r)/2\pi$ mod $n_G$. The
figure shows those changes.  The short solid lines mark the
non-zero changes of $n_G\th^{aa}(\v i)/2\pi$. The filled circles
mark the non-zero changes of $n_G\th^{xy}(\v i+\frac{\v
x}{2}+\frac{y}{2})/2\pi$ \etc. All the changes are equal to 1. }
\label{Qiix}
\end{figure}

In large $\t U_1$ and $\t U_2$ limit, the low energy Hilbert space
satisfy the scalar and vector constraints
\begin{eqnarray}
\label{constr1}
S(\v i)&\equiv &\exp[i\eta(\v i)]=1
\\
\label{constr2}
V(\v i,\v i+ \v a)&\equiv &\exp[2\pi i Q(\v i,\v i+\v a)/n_G]=1
\end{eqnarray}
where $\textbf{a}=\textbf{x},\textbf{y},\textbf{z}$.
The scalar constraint has the following form
\begin{equation}
\label{Scabr}
 S(\v i)=\exp(i\sum_{\v r} c^{ab}_{\v i,\v r} \th_{ab}(\v r))
\end{equation}
The structure of the integer coefficients $c^{ab}_{\v i,\v r}$ can
be seen more clearly in Fig. \ref{eta}. Fig. \ref{eta} also shows
the action of $S(\v i)$. The vector constraint has the following
form
\begin{equation}
\label{Vdabr}
 V(\v i,\v i+\v x)= \exp(i\sum_{\v r}
d^{ab}_{\v i,\v i+\v x,\v r} \frac{2\pi}{n_G}L^{ab}(\v r))
\end{equation}
The structure of the integer coefficients $d^{ab}_{\v i,\v i+\v
x,\v r}$ and the action of $V(\v i,\v i+\v x)$ are plotted in Fig.
\ref{Qiix}.

{}From the Figs. \ref{eta} and \ref{Qiix}, one can see that
$ \sum_{\v r}
c^{ab}_{\v i,\v r} d^{a'b'}_{\v j,\v j+\v x,\v r} =0$,
where $\sum_{\v r}$ sum over all the vertices and the square faces
of the cubic lattice. So the scalar and vector constraints commute
with each other,
$ [S(\v i),V(\v j,\v j+\v a)]=0$, $\v a =\v x, \v y, \v z$.
We can construct states that satisfy all the constraints one by one.

The states of the spins on vertices and squares can be labeled as
\begin{eqnarray}
|n\rangle_{\v i,ab} \quad n=0,1,2\cdots n_G-1
\end{eqnarray}
which satisfy $|n\rangle=|n+n_G\rangle$ and
\begin{eqnarray}
W^{ab}_L(\v i)|n\rangle_{\v i,ab}&=&e^{i2\pi n/n_G}|n\rangle_{\v
i,ab} \nonumber\\ W^{ab}_\theta(\v i)|n\rangle_{\v
i,ab}&=&|n+1\rangle_{\v i,ab}
\end{eqnarray}
it is easy to show $\left(W^{ab}_\theta(\v
i)\right)^{n_G}|n\rangle_{\v i,ab}=|n\rangle_{\v i,ab}$.

The simplest state that satisfy the vector constraints (\ref{constr2}) is the
state $|0\rangle=\bigotimes_{\v i, ab}|0\rangle_{\v i,ab}$.  However, this
state does not satisfy scalar constraint (\ref{constr1}). Note that $S(\v
i)|0\rangle$ is a new state still satisfies the vector constraints
(\ref{constr2}).  Since $\left(S(\v i)\right)^n|0\rangle$ satisfies
(\ref{constr2}) and $S(\v i)^{n_G}|0\rangle=|0\rangle$, therefore the state
$|\t 0\rangle=\bigotimes_{\v i}\left(\sum_{n=0}^{n_G} S(\v
i)^n\right)|0\rangle$.  satisfies both the vector constraints (\ref{constr1})
and scalar constraint (\ref{constr2}).  Similarly, starting from any state that
satisfies the vector constraints (\ref{constr2}), we can apply the operator $
\otimes_{\v i}\left(\sum_{n=0}^{n_G} S(\v i)^n\right)$ to obtain a state that
satisfies all the constraints.

Now we only need to construct the states that satisfy the vector
constraints (\ref{constr2}). All these states can be generated by
acting the operators $\cR_{xx}(\v i),\cR_{yy}(\v i),\cR_{zz}(\v
i),\cR_{xy}(\v i+\v x/2+\v y/2),\cR_{yz}(\v i+\v y/2+\v
z/2),\cR_{zx}(\v i+\v z/2+\v x/2)$ on vacuum $|0\rangle$. Those
operators are actually the discretized version of $R^{ij}$, for example:
\begin{eqnarray}
\label{cRyz} &&\cR_{yz}(\v i+\frac{\v y}{2}+ \frac{\v z
}{2})=\exp\{i[2\theta_{xx}(\v i)+2\theta_{xx}(\v i+\v
y)\nonumber\\&&+2\theta_{xx}(\v i+\v z)+2\theta_{xx}(\v i+\v y+\v
z)-\theta_{xy}(\v i+\frac{\v x}{2}+\frac{\v
y}{2})\nonumber\\&&-\theta_{xy}(\v i-\frac{\v x}{2}+\frac{\v y}{2})
-\theta_{zx}(\v i+\frac{\v x}{2}+\frac{\v z}{2})-\theta_{zx}(\v
i-\frac{\v x}{2}+\frac{\v z}{2})\nonumber\\&&-\theta_{xy}(\v i+\v
z+\frac{\v x}{2}+\frac{\v y}{2})-\theta_{xy}(\v i+\v z-\frac{\v
x}{2}+\frac{\v y}{2})\nonumber\\&&-\theta_{zx}(\v i+\v y+\frac{\v
x}{2}+\frac{\v z}{2})-\theta_{zx}(\v i+\v y-\frac{\v x}{2}+\frac{\v
z}{2})\nonumber\\&&+\theta_{yz}(\v i+\v x+\frac{\v y}{2}+\frac{\v
z}{2})+\theta_{yz}(\v i-\v x+\frac{\v y}{2}+\frac{\v
z}{2})\nonumber\\&&+2\theta_{yz}(\v i+\frac{\v y}{2}+\frac{\v
z}{2})]\}
\end{eqnarray}

\begin{eqnarray}
\label{cRzz} &&\cR_{zz}(\v i)=\exp\{i[\theta_{xx}(\v i+\v
y)+\theta_{xx}(\v i-\v y)+2\theta_{xx}(\v
i)\nonumber\\&&+\theta_{yy}(\v i+\v x)+\theta_{yy}(\v i-\v
x)+2\theta_{yy}(\v i)-\theta_{xy}(\v i+\frac{\v x}{2}+\frac{\v
y}{2})\nonumber\\&&-\theta_{xy}(\v i-\frac{\v x}{2}+\frac{\v y}{2})
-\theta_{xy}(\v i+\frac{\v x}{2}-\frac{\v y}{2})-\theta_{xy}(\v
i-\frac{\v x}{2}-\frac{\v y}{2})]\}\nonumber\\
\end{eqnarray}

Other components can be easily obtained by cycling $xyz$ to $yzx$
and $zxy$.

\begin{figure}[tbp]
\begin{center}
\includegraphics[scale=0.5] {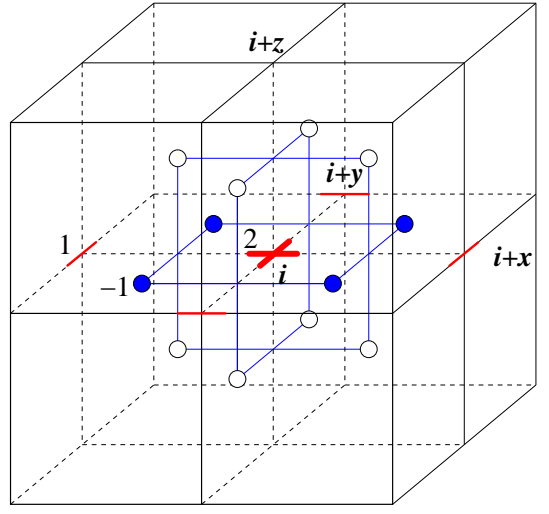}
\end{center}
\caption{
(Color online)
The action of the operator $\cR_{zz}(\v i)$ in \eqn{cRzz}
changes $L^{ab}(\v r)$ mod $n_G$.
The short solid lines mark the non-zero changes of $L^{aa}(\v i)$.
The thicker lines
correspond to an change of 2 and thinner lines a change of 1.
The filled circles mark the non-zero
changes of $L^{xy}(\v i+\frac{\v x}{2}+\frac{y}{2})$
\etc. The changes are equal to $-1$.
} \label{Rzz}
\end{figure}

\begin{figure}[tbp]
\begin{center}
\includegraphics[scale=0.5] {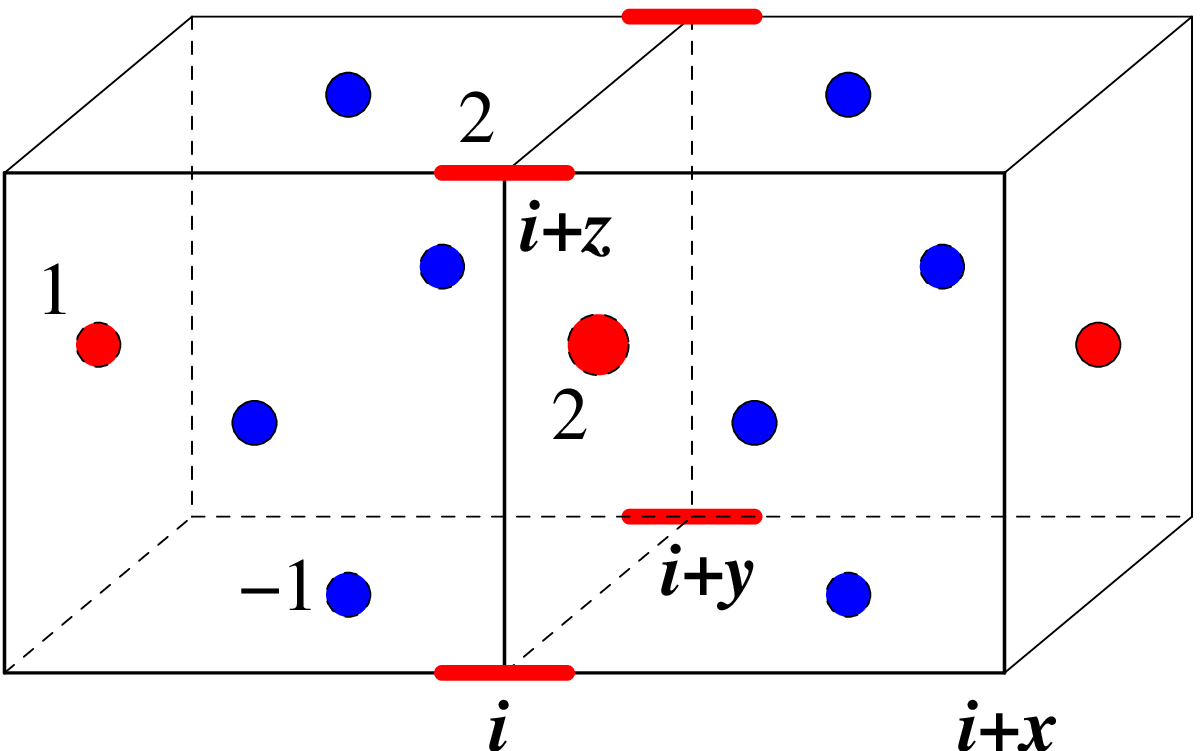}
\end{center}
\caption{
(Color online)
The action of the operator $\cR_{yz}(\v i+\frac{\v y}{2}+\frac{\v z}{2})$
in \eqn{cRyz}
changes $L^{ab}(\v r)$ mod $n_G$.
The short solid lines mark the non-zero changes of $L^{aa}(\v i)$
where the changes are equal to $2$.
The filled circles mark the non-zero
changes of $L^{xy}(\v i+\frac{\v x}{2}+\frac{y}{2})$
\etc,  where the bigger circle corresponds to a change of $2$
and smaller circles a change of $\pm 1$.
Red represents positive integers and blue negative integers.
} \label{Ryz}
\end{figure}

Figures \ref{Rzz} and \ref{Ryz} illustrate the action of $\cR_{ab}$.
Comparing the patterns in Figs. \ref{Rzz} and \ref{Ryz} with the
pattern in Fig. \ref{Qiix}, we can show that the operators
$\cR_{ab}$ commute with the vector constraint $V(\v i,\v i+\v a)$.
As functions of $\th_{ab}$, the operators $\cR_{ab}$ also commute
with the scalar constraint $S(\v i)$.  The operators that commute
with the scalar and the vector constraints are called gauge
invariant operators.  It is easy to check $\cR_{ab}$ are the
\emph{complete} gauge invariant operators and the low energy Hilbert
space can be constructed by acting these operators on $|\t 0\rangle$
one or more times. But one should notice not all such operations can
successfully create new states, for example, $\cR_{xx}(\v
i)\cR_{yy}(\v i)\cR_{zz}(\v i)|\t 0\rangle=S(\v i)|\t 0\rangle=|\t
0\rangle$.

\section{The L-type lattice model}

\kp{\item
Giving non-trivial dynamics to states in
the $H_U=0$ subspace through
$H^L_J$ and $H^L_g$.
}

In the last section, we have considered spin model whose Hilbert
space is defined through the algebra \eq{thLalg} and whose
Hamiltonian is given by $H_U$.  We find that such a system has
infinite many zero-energy states and all other states have an energy
at least of order $\t U_{1,2}/n_G$.  In this section, we are going
to add additional terms to the Hamiltonian that act within the
zero-energy subspace.  The zero-energy subspace is called the
constrained subspace and the operators that act within the
constrained subspace are called gauge invariant operators. The new
term will lift the degeneracy of the zero-energy states and give
those states a non-trivial dynamics.  We find that the resulting low
energy collective modes contain only helicity $\pm 2$ modes with
$\om \sim k^3$ dispersion.

\subsection{Putting the $R^{ij}R^{ij}$ term on lattice}

We have seen that $\cR^{ab}$ is a gauge invariant operator
which acts within the constrained subspace.
So we can add the following term
\begin{align*}
H_g^L&=-\frac{n_G g^L}{2}\left[\sum_{\v i, a=\v x,\v y,\v
z}\cR_{aa}+\sum_{\v i, ab=\v xy,\v yz,\v zx}\frac{\cR_{ab}}{2}+h.c.
\right]
\end{align*}
to the Hamiltonian.  It turns out that if we expand gauge invariant
operator $\cR^{ab}$ to quadratic order in $\th^{ij}$, we will obtain
the gauge invariant term $R^{ij}R^{ij}$.  Thus the above lattice
Hamiltonian term corresponds the $R^{ij}R^{ij}$ term in the
continuum model.

\subsection{Putting the $C^i_jC^i_j$ term on lattice}

To put the  $C^i_jC^i_j$ term in the continuum model on lattice,
we first put the field $C^i_j$ on lattice. From the relation
between continuum field $\cE^{ij}$ and the lattice operator
$L^{ij}$, we find a set of lattice operators $\si^i_j$ that
correspond to $C^i_j$. $\si^i_j$ are given by, for example
\begin{eqnarray}
\label{eqsixx}
\sigma^x_x&=&-L^{zx}(\v i+\frac{\v x}{2}+\frac{\v z}{2})-L^{zx}(\v
i+\frac{\v x}{2}+\frac{\v z}{2}+\v y)\nonumber\\&+&L^{xy}(\v
i+\frac{\v x}{2}+\frac{\v y}{2})+L^{xy}(\v i+\frac{\v
x}{2}+\frac{\v y}{2}+\v z)
\end{eqnarray}

\begin{eqnarray}
\label{eqsixy}
\sigma^x_y&=&
 2L^{yz}(\v i+\frac{\v y}{2}+\frac{\v z}{2})
+2L^{yz}(\v i-\frac{\v y}{2}+\frac{\v z}{2})\nonumber\\
&-&\left[L^{yy}(\v i+\v z)-L^{xx}(\v i+\v z)-L^{zz}(\v i+\v z)\right]
\nonumber\\
&-&\left[L^{yy}(\v i)-L^{xx}(\v i)-L^{zz}(\v i)\right]
\end{eqnarray}

\begin{eqnarray}
\label{eqsixz}
\sigma^x_z&=&
\left[L^{zz}(\v i+\v y)-L^{xx}(\v i+\v y)-L^{yy}(\v i+\v y)\right]
\nonumber\\
&+&\left[L^{zz}(\v i)-L^{xx}(\v i)-L^{yy}(\v i)\right]
\nonumber\\
&-& 2L^{yz}(\v i+\frac{\v y}{2}+\frac{\v z}{2})
-2 L^{yz}(\v i+\frac{\v y}{2}-\frac{\v z}{2})
\end{eqnarray}

\begin{figure}[tbp]
\begin{center}
\includegraphics[scale=0.45] {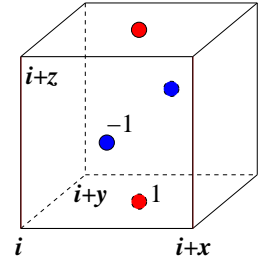}
\end{center}
\caption{
(Color online)
The integer coefficients $ f^x_{x,cd,\v r}$ in \eqn{sif}
(see \eqn{eqsixx}). The filled circles mark the non-zero $
f^x_{x,cd,\v r}$ \etc, which have values $\pm 1$. Red represents
positive integers and blue negative integers. } \label{sixx}
\end{figure}

\begin{figure}[tbp]
\begin{center}
\includegraphics[scale=0.45] {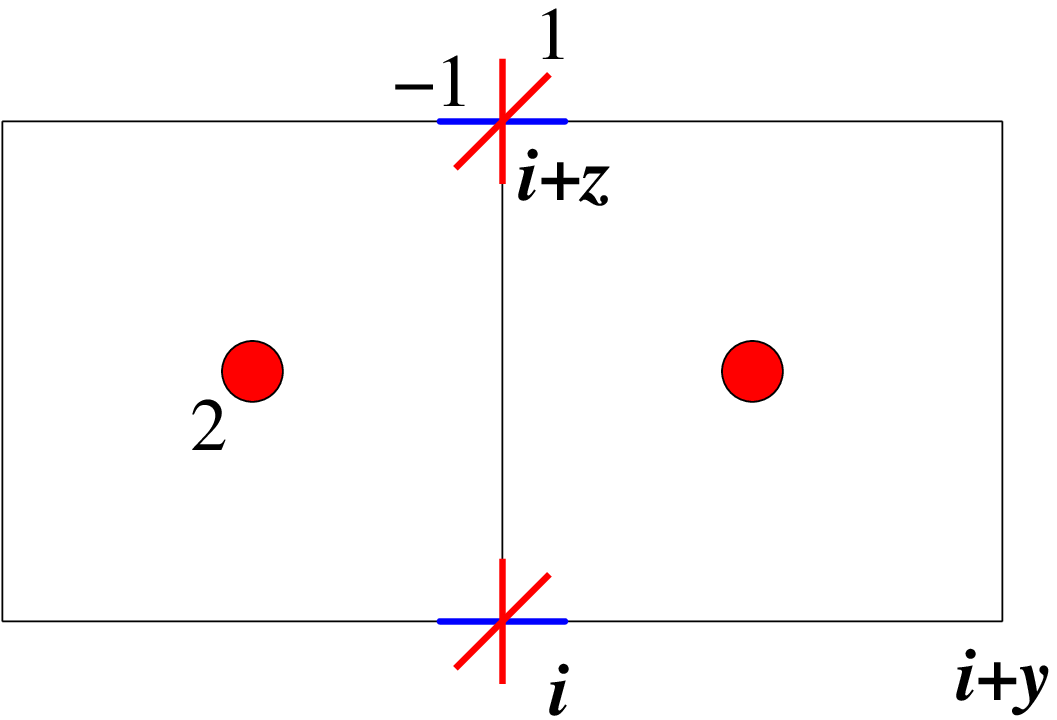}
\end{center}
\caption{
(Color online)
The integer coefficients $ f^y_{x,cd,\v r}$ in \eqn{sif}
(see \eqn{eqsixy}).  The short solid lines mark the non-zero
$f^y_{x,cc,\v j}$, which have values $\pm 1$. The filled circles
mark the non-zero $ f^y_{x,cd,\v r}$ \etc, which have values $2$.
Red represents positive integers and blue negative integers. }
\label{sixy}
\end{figure}

\begin{figure}[tbp]
\begin{center}
\includegraphics[scale=0.45] {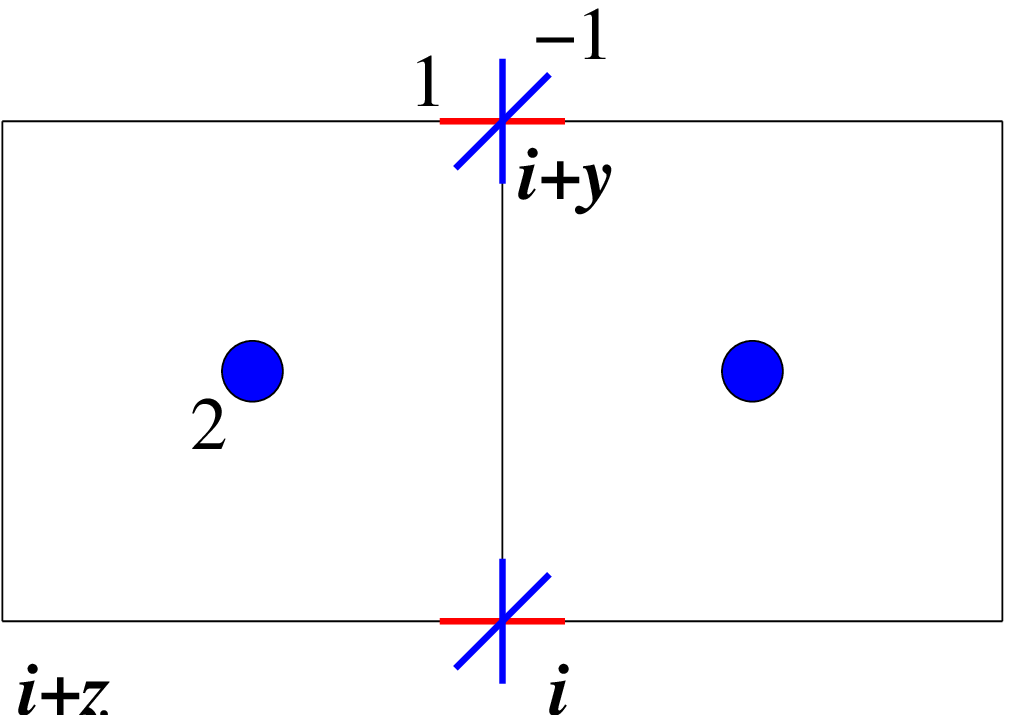}
\end{center}
\caption{
(Color online)
The integer coefficients $ f^x_{z,cd,\v r}$ in \eqn{sif}
(see \eqn{eqsixz}).  The short solid lines mark the non-zero
$f^x_{z,cc,\v j}$, which have values $\pm 1$. The filled circles
mark the non-zero $ f^x_{z,cd,\v r}$ \etc, which have values $-2$.
Red represents positive integers and blue negative integers. }
\label{sixz}
\end{figure}

We note that $\si^a_b$ have a form
\begin{equation}
\label{sif}
 \si^a_b=\sum_{c,d,\v r} f^a_{b,cd,\v r} L^{cd}(\v r)
\end{equation}
where $f^a_{b,cd,\v r}$ are integer coefficients.  Figs.
\ref{sixx}, \ref{sixy}, and \ref{sixz} show that pattern of those
coefficients.  From those patterns, one can check that the
$\si^i_j$ operators commute with the constraint operators $S(\v
i)$ and $V(\v i,\v i+\v a)$.

After compactification, the corresponding physical operators are
defined as
\begin{eqnarray}
T_a^b=\exp\left[\frac{2\pi i}{n_G}\sigma_a^b\right]
\end{eqnarray}
Thus Hamiltonian term $C^i_jC^i_j$ in the continuum model
become the follow lattice Hamiltonian term
\begin{align*}
H_J^L&=-n_GJ^L\left(
 \sum_{\v i, a=x,y, z} \frac{T_a^a}{2}
+\sum_{\v i, ab= xy, yz, zx}\frac{T_a^b+T_b^a}{8}
+h.c. \right)
\end{align*}

Putting everything together, we obtain the following
lattice Hamiltonian
\begin{eqnarray}
\label{HUJgL}
H&=&H_U+H_J^L+H_g^L
\end{eqnarray}
Such a lattice Hamiltonian defines a L-type lattice model.

\subsection{Low energy collective modes
-- spin-wave/quantum-freeze approach}

\label{spinwv}

\kp{\item
The only gapless exactions in the L-type model
are helicity $\pm 2$ modes with $\om \sim k^3$ dispersion.
\item
The low energy effective theory for the L-type model has an
emergent diffeomorphism gauge symmetry.
}

In this section, we will use spin-wave/quantum-freeze approach to
study the low energy dynamics of the L-type model \eq{HUJgL}. Another
approach, the equation-of-motion approach, will be given in appendix
\ref{eomLtype}.

In the spin-wave/quantum-freeze approach, we first assume the quantum
fluctuations of $\th_{ a b}$ and $\phi^{ a b}=2\pi L^{ab}/n_G$ to be much
larger than their discreteness $\Del \th_{ a b}\sim \Del \phi^{ a b} \sim
1/n_G$ in the large $n_G$ limit. (We will check this assumption later for self
consistency.) In this case, we can treat $\th_{ a b}$ and $\phi^{ a b}$ as
classical fields and use the classical phase-space Lagrangian \eq{Lthphi} to
study the low energy collective modes.

\subsubsection{Classical spin waves}

Let us also assume that  the quantum fluctuations of $\th_{a b}$,
$\phi^{ab}$ are much smaller then 1 so that we can expand the phase-space
Lagrangian to quadratic order in  $\th_{ab}$, $\phi^{ab}$.
This allows us
to find the dispersions of collective modes of our bosonic model
\eq{HUJgL}.  There are total of six collective modes. We find four of
them have zero frequency for all $\v k$, and two modes have a cubic
dispersion relation near $\v k=(\pi,\pi,\pi)$. Near $\v
k=(\pi,\pi,\pi)$ , the dynamics of the six modes are described by
the following continuum field theory:
\begin{align}
\label{Lagthphi}
\cL & =n_G\Big[ \frac{\phi^{ij}}{2\pi} \dot
\theta_{ij} -\frac{ J^L}{2}C_j^i C_j^i  -\frac{
g^L}{2}R^{ij}R^{ij}\Big]
\nonumber\\
&\ \ \ \ \ \ \ \ \ \ -\frac{n_G\t U_1}{2}(\prt_i\phi^{ij})^2
-\frac{n_G\t U_2}{2} (R^{ii})^2
\end{align}
where $R^{ij}=\eps^{imk}\eps^{jln}\prt_m\prt_l \theta_{nk}$ and
$C^i_j=\epsilon^{imn}\partial_m\left(\phi^{nj}-\frac{1}{2}\delta_{nj}\phi^{ll}\right)$.
The continuum fields $\theta_{ab}(\v x)$ are given by $-(-1)^{\v
i}\frac12 \theta^{ab}(\v i+\frac{\v a}{2}+\frac{\v b}{2})$ for
$a\neq b$ and by $(-1)^{\v i} \theta^{ab}(\v i)$ for $a=b$, where
$a,b=x,y,z$. The continuum fields $\phi^{ab}(\v x)$ are given by
$-(-1)^{\v i} \phi^{ab}(\v i+\frac{\v a}{2}+\frac{\v b}{2})$ for
$a\neq b$ and by $(-1)^{\v i} \phi^{ab}(\v i)$ for $a=b$.

\subsubsection{The classical helicity $\pm 2$ modes}

The two modes that have  a cubic dispersion relation are the two
helicity $\pm 2$ modes. The dispersion relation is given by $\om_{\v
k} \sim \sqrt{ g^L J^L} |\v k|^3$.  We note that the $\t U_2$ and
$\t U_1$ terms decouple from the helicity $\pm 2$ modes, since the
helicity $\pm 2$ modes satisfy the constraints
$\prt_i\phi^{ij}=R^{ii}=0$. Thus the dynamics of the helicity $\pm
2$ modes does not depend on $\t U_2$ and $\t U_1$.

To see if the result $\om_{\v k} \sim \sqrt{ g^L J^L} |\v k|^3$ can
be trusted, we need to analyze the quantum fluctuations of
$\theta_{ij}$ and $\phi^{ij}$.  We find that for large $n_G$, the
quantum fluctuations of the helicity $\pm 2$ modes is of order $\del
\phi^{ij},\del \theta_{ij}\sim \sqrt{1/n_G}$ (note that $ J^L$ and $
g^L$ are of the same order).  So the fluctuations of $\phi^{ij}$ and
$\theta_{ij}$ satisfy $1/n_G\ll \del \phi^{ij}\ll 1$, $1/n_G\ll \del
\theta_{ij}\ll 1$ and the semiclassical approximation is valid for
the helicity $\pm 2$ modes. In this case, the result $\om_{\v k}\sim
\sqrt{\t g\t J}|\v k|^3$ can be trusted.

\subsubsection{The quantum freeze and the gapping of helicity $0$ and $\pm 1$
modes}

The helicity $\pm 1$ modes and one of the helicity $0$ mode are
described by $(\th_i,\phi_i)$ which correspond to fluctuations  of
the following form: $\theta_{ij}= \prt_i \th_j+\prt_j \th_i$ and
$\phi^i=\prt_j \phi^{ji}$.  Their frequency $\om_{\v k}=0$.  For
such modes, the Hamiltonian only contains $\phi^i$. Thus the quantum
fluctuations satisfies $\del \phi^{i} \ll 1/n_G$ and $\del
\th_{i}\gg 1$. So the semiclassical approximation is not valid and
the result $\om_{\v k}=0$ cannot be trusted.  Using the similar
argument used in the emergence of $U(1)$ gauge bosons in section
\ref{clssU1}, we conclude that those modes are quantum frozen and
are gapped.  Since discreteness of $\phi^{ij}$ is of order $1/n_G$,
the gap of the helicity $0$ and $\pm 1$ modes is of order $\t
U_1/n_G$.  The strong fluctuations $\del \theta_{ij}= \prt_i
\th_j+\prt_j \th_i \gg 1$ lead to gauge transformations
\begin{equation}
\label{G1}
 \theta_{ij}\to \theta_{ij}+ \prt_i \th_j+\prt_j \th_i
\end{equation}
and the weak fluctuations $\phi^i\ll 1/n_G$ lead to constraints
\begin{equation}
\label{C1}
 \prt_j \phi^{ji}=0
\end{equation}

The second  helicity $0$ mode is described by $(\th,\phi)$ which
correspond to the following fluctuations: $\phi^{ij}=
(\del_{ij}\prt^2-\prt_i\prt_j) \phi$ and $\th=
(\del_{ij}\prt^2-\prt_i\prt_j)\theta_{ij}$.  Its frequency is again
$\om_{\v k}=0$. The Hamiltonian for such a mode contains only $\th$.
So the quantum fluctuations satisfies $ \del \phi \gg 1$ and $\del
\th\ll 1/n_G$. The second helicity $0$ mode is also gapped with a
gap of order $\t U_2$.  The strong fluctuations $\del \phi^{ij}=
(\del_{ij}\prt^2-\prt_i\prt_j) \phi \gg 1$ lead to a gauge
transformation
\begin{equation}
\label{G2}
\phi^{ij}\to \phi^{ij}+(\del_{ij}\prt^2-\prt_i\prt_j) \phi
\end{equation}
and the weak fluctuations $ \th=
(\del_{ij}\prt^2-\prt_i\prt_j)\theta_{ij} \ll 1/n_G$ lead to a
constraint
\begin{equation}
\label{C2} (\del_{ij}\prt^2-\prt_i\prt_j)\theta_{ij}=0
\end{equation}

The key in our argument is the gapping of helicity $0$ and $\pm 1$
modes. Clearly, in the L-type lattice model \eq{HUJg}, those modes
are strongly fluctuating and strongly interacting modes. Those modes
also have a very narrow band width and arise from compact degree
freedom on lattice.  The strongly interacting modes with flat bands
are in general gapped.  The gapless helicity $\pm 2$ modes are very
classical in the large $n_G$ limit. They should survive the gapping
of the  helicity $0$ and $\pm 1$ modes.

\section{The N-type lattice model}

\kp{\item
Giving non-trivial dynamics to states in
the low energy Hilbert space through $H_J$ and $H_g$.
}

In this section, we will try to construct a lattice model whose low
energy effective theory is the N-type model \eq{LNtype}  with the
constraints \eqn{T1} and \eqn{T9}.\cite{GW0600} Just like the L-type
lattice model, the Hamiltonian of the N-type lattice model contains
$H_U$ \eq{HUC} which imposes the constraints \eqn{T1} and \eqn{T9}.
Other terms of the Hamiltonian can be obtained by putting the
continuum Hamiltonian \eq{T17} on lattice.

\subsection{Putting $a_{ij}R^{ij}$ on lattice}

If we introduce
$D^l_k = \eps^{ijl} \prt_i a_{jk}$,
we find that
$ a_{ij}R^{ij}\sim D^l_k D^k_l$.
On lattice $D^l_k$ can be chosen to have the following form
\begin{align*}
  D^x_x &=\prt_y a_{zx}-\prt_z a_{yx}\sim \frac12 \rho^x_x,
\nonumber\\
  D^x_y &=\prt_y a_{zy}-\prt_z a_{yy}\sim \frac12 \rho^x_y,
\nonumber\\
  D^x_z &=\prt_y a_{zz}-\prt_z a_{yz}\sim \frac12 \rho^x_z ,
\end{align*}
where
Here $\rho^i_j(\v i)$ are defined as
\begin{align*}
\rho^x_x(\v i) & =
  \th_{zx}(\v i+\v y +\frac{\v z}{2}+\frac{\v x}{2})
+ \th_{zx}(\v i+\frac{\v z}{2}+\frac{\v x}{2})
\nonumber\\ &
- \th_{xy}(\v i+\v z +\frac{\v x}{2}+\frac{\v y}{2})
- \th_{xy}(\v i +\frac{\v x}{2}+\frac{\v y}{2}) ,
\nonumber\\
\rho^x_y(\v i) & =
- \th_{zy}(\v i +\frac{\v z}{2}+\frac{\v y}{2})
- \th_{zy}(\v i+\frac{\v z}{2}-\frac{\v y}{2})
\nonumber\\ &
+2\th_{yy}(\v i+\v z)
+2\th_{yy}(\v i) ,
\nonumber\\
\rho^x_z(\v i) & =
-2\th_{zz}(\v i+\v y)
-2\th_{zz}(\v i)
\nonumber\\ &
+ \th_{yz}(\v i+\frac{\v y}{2}+\frac{\v z}{2})
+ \th_{yz}(\v i+\frac{\v y}{2}-\frac{\v z}{2}) .
\end{align*}
Other components are obtained by cycling $xyz$ to $yzx$ and $zxy$.
So $\frac 12 g a_{ij}R^{ij}$ leads to  the following contribution to the
Hamiltonian
\begin{align*}
 H_g' &= \frac18 g\sum_{\v i,a=x,y,z} [\rho^a_a(\v i)]^2
 + \frac14 g\sum_{\v i,ab=xy,yz,zx} \rho^a_b(\v i)\rho^b_a(\v i)
\end{align*}

In the compactified model, the following term
\begin{align*}
 H_g &= \frac{n_G \t g}{4}\sum_{\v i,a=x,y,z} \{1-\cos[\rho^a_a(\v i)]\}
\nonumber\\ &
 + \frac{n_G \t g}{4}\sum_{\v i,ab=xy,yz,zx} \sin[\rho^a_b(\v i)]\sin[\rho^b_a(\v i)]
\end{align*}
reproduce $H_g'$ in the small $\rho^a_b$ limit.  The operators in $H_g$ are
functions of $W_L^{ab}$ and $W_\th^{ab}$. So $H_g$ acts within the Hilbert
space of compactified model.

\subsection{Putting $\cE^{ij}\cE^{ij}-\frac12 \cE^{ii}$ on lattice}

The term $\frac12 J[\cE^{ij}\cE^{ij}-\frac12 \cE^{ii}]$
can be put on lattice easily. It leads to the following contribution to the
Hamiltonian
\begin{align*}
 H_J'&=J
\sum_{\v i,ab=xy,yz,zx} L^{ab}(\v i+\frac{\v a}{2}+\frac{\v b}{2})L^{ab}(\v
i+\frac{\v a}{2}+\frac{\v b}{2})
\nonumber\\
&\ \ \ -\frac12 J \sum_{\v i,a=x,y,z} [L^{aa}(\v i)]^2
\end{align*}
In the compactified model, the following term
\begin{align*}
  H_J &= n_G \t J\sum_{\v i,a=x,y,z} \{1-\cos[2\pi L^{aa}(\v i)/n_G]\}
\nonumber\\ &
  + 2n_G \t J\sum_{\v i}\sum_{ab=xy,yz,zx} \{1-\cos[2\pi L^{ab}(\v i)/n_G]\}
\nonumber\\ &
  - \frac12 n_G \t J\sum_{\v i}\{1-\cos[2\pi \sum_{a=x,y,z} L^{aa}(\v i)/n_G]\}
\end{align*}
reproduce $H_J'$ in the small $L^{ab}/n_G$ limit.\\

\subsection{The N-type lattice model}

Putting everything together, we obtain the following lattice
Hamiltonian
\begin{align}
\label{HUJg} H &= H_U+H_J+H_g
\end{align}
which defines the N-type lattice spin model.  We will assume $U\sim J
\sim g$ in the N-type lattice model.
The corresponding phase-space Lagrangian is given by
\begin{align}
\label{Lthphi}
 L &=\frac{n_G}{2\pi}
\sum_{\v i,ab=xy,yz,zx}\phi^{ab}(\v i+\frac{\v a}{2}+\frac{\v b}{2})
\prt_0 \th_{ab}(\v i+\frac{\v a}{2}+\frac{\v b}{2})
\nonumber\\ &
+\frac{n_G}{2\pi}\sum_{\v i,a=x,y,z} \phi^{aa}(\v i)\prt_0 \th_{aa}(\v i)
-H(\phi^{ij},\th_{ij})
\end{align}
where $\phi^{ij}=2\pi L^{ij}/n_G$
and $H(\phi^{ij},\th_{ij})$ is given by \eq{HUJg} with $L^{ij}$
replaced by $n_G\phi^{ij}/2\pi$.

\subsection{Low energy collective modes
-- spin-wave/quantum-freeze approach}

\label{spinwvN}

\kp{\item
In a semiclassical approach to the N-type model,
we can expand the classical action to the quadratic
order and use the spin-wave/quantum-freeze approach
to obtain the low energy dynamics.}

The motivation of constructing the quantum spin model \eq{HUJg} or \eq{Lthphi}
is to obtain a lattice model whose low lying collective modes are gravitational
waves with a linear dispersion.  After we obtain the model \eq{HUJg}, we throw
always the motivation and ask ``do the low energy excitations of the spin model
\eq{HUJg} really correspond to gravitons?''

The spin model \eq{HUJg} is a complicated and strongly interacting quantum
system. It seems impossible to obtain and to understand the dynamics of its low
energy excitations.  On the other hand, the rotor model \eq{lattLC} is also a
complicated and strongly interacting quantum system. Many different approaches,
such as equation-of-motion approach, coherent state approach, and
spin-wave/quantum-freeze approach, are developed to understand the low energy
dynamics of \eq{lattLC}.\cite{Walight,Wen04,Wqoem,LWqed}

In this section, we will use the spin-wave/quantum-freeze approach developed in
section \ref{clssU1} to understand the low energy excitations of the spin
system described by the Hamiltonian \eq{HUJg} in large $n_G$ limit with $\t g$,
$\t J$, $\t U_1$ and $\t U_2$ are of the same order.  We would like to remark
that the spin-wave/quantum-freeze approach has not been tested fully and may
not be as reliable as the equation-of-motion approach used in section
\ref{eomLtype}.  We note that for the N-type lattice model, $H_J$ and $H_g$ do
not act within the zero-energy subspace of $H_U$ (since $[H_U,H_J]\neq 0$, and
$[H_U,H_g]\neq 0$).  The equation-of-motion approach cannot be used for the
N-type lattice model.  This is why we will use spin-wave/quantum-freeze
approach to analyze the low energy excitations of the N-type lattice model. The
results obtained in this section may not be reliable and need to be confirmed
through other more reliable methods.

In the spin-wave/quantum-freeze approach, we first assume
the quantum fluctuations of $\th_{\v i\v j}$ and $\phi_{\v i\v j}$
to be much larger than their discreteness $\Del \th_{\v i\v j}\sim
\Del \phi_{\v i\v j} \sim 1/n_G$. (We will check this assumption
later for self consistency.) In this case, we can treat $\th_{\v
i\v j}$ and $\phi_{\v i\v j}$ as classical fields and use the
classical phase-space Lagrangian \eq{Lthphi} to study the low
energy collective modes.

\subsubsection{Classical spin waves}

Let us also assume that  the quantum fluctuations of $\th_{\v i\v j}$
and $\phi_{\v i\v j}$ are much smaller then 1 and expand the
phase-space Lagrangian to quadratic order in  $\th_{\v i\v j}$
and $\phi_{\v i\v j}$.
In $\v k$-space, the quadratic Lagrangian has a form:
\begin{widetext}
\begin{eqnarray}
&& \mathcal{L}=\frac{n_G}{2\pi}\sum_{\v{k}} \left[\begin{array}{c}\phi^{xx}(-\v{k})\nonumber\\
\phi^{yy}(-\v{k})
\nonumber\\ \phi^{zz}(-\v{k})\nonumber\\
\phi^{xy}(-\v{k})
\nonumber\\ \phi^{yz}(-\v{k})\nonumber\\
\phi^{zx}(-\v{k})
\end{array}\right]^t
\left[\begin{array}{cccccc}1 & 0& 0& 0& 0& 0\nonumber\\0 & 1& 0& 0&
0& 0\nonumber\\0 & 0& 1& 0& 0& 0\nonumber\\0 & 0& 0& 1& 0&
0\nonumber\\0 & 0& 0& 0& 1& 0\nonumber\\0 & 0& 0& 0& 0& 1
\end{array}\right]\left[\begin{array}{c}\dot{\theta}^{xx}(\v{k})\nonumber\\\dot{\theta}^{yy}(\v{k})
\nonumber\\ \dot{\theta}^{zz}(\v{k})\nonumber\\
\dot{\theta}^{xy}(\v{k})
\nonumber\\ \dot{\theta}^{yz}(\v{k})\nonumber\\
\dot{\theta}^{zx}(\v{k})
\end{array}\right]-n_G \sum_{\v{k}}
\left[\begin{array}{c}\phi^{xx}(-\v{k})\nonumber\\
\phi^{yy}(-\v{k})
\nonumber\\ \phi^{zz}(-\v{k})\nonumber\\
\phi^{xy}(-\v{k})
\nonumber\\ \phi^{yz}(-\v{k})\nonumber\\
\phi^{zx}(-\v{k})
\end{array}\right]^t \left\{
\widetilde{J}\left[\begin{array}{cccccc}\frac{1}{4} &-\frac{1}{4} &
-\frac{1}{4}& 0& 0& 0\nonumber\\-\frac{1}{4} & \frac{1}{4}&
-\frac{1}{4}& 0& 0& 0\nonumber\\-\frac{1}{4} & -\frac{1}{4}&
\frac{1}{4}& 0& 0& 0\nonumber\\0 & 0& 0& 1& 0& 0\nonumber\\0 & 0& 0&
0& 1& 0\nonumber\\0 & 0& 0& 0& 0& 1
\end{array}\right]
\right.
\nonumber\\ &&
\left.
+2\widetilde{U}_2
\left[\begin{array}{cccccc}c_x^2
& 0& 0& c_{xy} & 0&
c_{zx}\nonumber\\0 &
c_y^2& 0& c_{xy} &
c_{yz} & 0\nonumber\\0 & 0&
c_z^2& 0& c_{yz} &
c_{zx}\nonumber\\
c_{xy}  &
c_{xy} & 0&
c_x^2+c_y^2 &
c_{zx}&
c_{yz} \nonumber\\0 &
c_{yz} &
c_{yz} &
c_{zx}&
c_y^2+c_z^2&
c_{xy} \nonumber\\
c_{zx} & 0&
c_{zx}&
c_{yz} &
c_{xy} &
c_z^2+c_x^2
\end{array}\right]\right\}\left[\begin{array}{c}\phi^{xx}(\v{k})\nonumber\\ \phi^{yy}(\v{k})
\nonumber\\ \phi^{zz}(\v{k})\nonumber\\
\phi^{xy}(\v{k})
\nonumber\\ \phi^{yz}(\v{k})\nonumber\\
\phi^{zx}(\v{k})
\end{array}\right]
+
\left[\begin{array}{c}\theta^{xx}(-\v{k})\nonumber\\
\theta^{yy}(-\v{k})
\nonumber\\ \theta^{zz}(-\v{k})\nonumber\\
\theta^{xy}(-\v{k})
\nonumber\\ \theta^{yz}(-\v{k})\nonumber\\
\theta^{zx}(-\v{k})
\end{array}\right]^t \left\{
\widetilde{g}\left[\begin{array}{cccccc}0 &-2c_z^2 &
-2c_y^2& 0& 2c_{yz} &
0\nonumber\\-2c_z^2 & 0& -2c_x^2& 0& 0&
2c_{zx}\nonumber\\-2c_y^2
& -2c_x^2& 0 & 2c_{xy} & 0&
0\nonumber\\0 & 0& 2c_{xy} &
c_z^2& -c_{zx}&
-c_{yz} \nonumber\\
2c_{yz}  & 0& 0&
-c_{zx}& c_x^2&
-c_{xy} \nonumber\\0 &
2c_{zx}& 0&
-c_{yz} &
-c_{xy} & c_y^2
\end{array}\right]\right.
\nonumber\\&&
\ \ \ \ \ \ \ \ \ \ \ \ \ \ \ \ \ \
+8\widetilde{U}_1\left[\begin{array}{cccccc}c_y^2+c_z^2
& 0& 0& 0& 0& 0\nonumber\\0 &
c_z^2+c_x^2& 0& 0& 0& 0\nonumber\\0 & 0&
c_x^2+c_y^2& 0& 0& 0\nonumber\\0 & 0& 0&
c_{xy} & 0& 0\nonumber\\0 & 0& 0& 0&
c_{yz} & 0\nonumber\\0 & 0& 0& 0& 0&
c_{zx}
\end{array}\right]
\left[\begin{array}{cccccc}1 & 1& 1& -1& -1&
-1\nonumber\\1 & 1& 1& -1& -1& -1\nonumber\\1 & 1& 1& -1& -1&
-1\nonumber\\-1 & -1& -1& 1& 1& 1\nonumber\\-1 & -1& -1& 1& 1&
1\nonumber\\-1 & -1& -1& 1& 1& 1
\end{array}\right]
\times\nonumber\\&&
\left.
\ \ \ \ \ \ \ \ \ \ \ \ \ \ \ \ \ \ \ \ \ \ \ \ \ \
\left[\begin{array}{cccccc}c_y^2+c_z^2
& 0& 0& 0& 0& 0\nonumber\\0 &
c_z^2+c_x^2& 0& 0& 0& 0\nonumber\\0 & 0&
c_x^2+c_y^2& 0& 0& 0\nonumber\\0 & 0& 0&
c_{xy} & 0& 0\nonumber\\0 & 0& 0& 0&
c_{yz} & 0\nonumber\\0 & 0& 0& 0& 0&
c_{zx}
\end{array}\right] \right\}
\left[\begin{array}{c}\th^{xx}(\v{k})\nonumber\\
\th^{yy}(\v{k})
\nonumber\\ \th^{zz}(\v{k})\nonumber\\
\th^{xy}(\v{k})
\nonumber\\ \th^{yz}(\v{k})\nonumber\\
\th^{zx}(\v{k})
\end{array}\right]
\end{eqnarray}
\end{widetext}
where $c_{x,y,z}$ and $c_{xy,yz,zx}$ are defined as, for example $
c_x=\cos\frac{k_x}{2},\ c_{xy}= \cos\frac{k_x}{2}\cos\frac{k_y}{2} $

After solving the equation of motions obtained from the quadratic Lagrangian,
we can find the dispersions of collective modes of our bosonic model
\eq{HUJg}.  There are total of six collective modes. We find four of them have
zero frequency for all $\v k$, and two modes have a linear dispersion relation
near $\v k=(\pi,\pi,\pi)$. Near $\v k=(\pi,\pi,\pi)$, the dynamics of the six
modes are described by the following continuum field theory:
\begin{align}
\label{LagthphiN}
\cL & =n_G\Big\{
\frac{\phi^{ij}}{2\pi} \dot a_{ij}
-\frac{\t J}{2}\Big[(\phi^{ij})^2 -\frac{(\phi^{ii})^2}{2}\Big]
-\frac{\t g}{2} \th_{ij}R^{ij}\Big\}
\nonumber\\
&\ \ \ \ \ \ \ \ \ \ -\frac{n_G\t U_1}{2}(\prt_i\phi^{ij})^2
-\frac{n_G\t U_2}{2} (R^{ii})^2
\end{align}
where $R^{ij}=\eps^{imk}\eps^{jln}\prt_m\prt_l a_{nk}$.
The
continuum fields $a_{bc}(\v x)$ are given by
$-(-1)^{\v i}\frac12 \theta^{bc}(\v
i+\frac{\v b}{2}+\frac{\v c}{2})$ for $b\neq c$ and by $(-1)^{\v i}
\theta^{bc}(\v i)$ for $b=c$, where $b,c=x,y,z$.
The
continuum fields $\phi^{ab}(\v x)$ are given by
$-(-1)^{\v i} \phi^{ab}(\v i+\frac{\v a}{2}+\frac{\v b}{2})$ for $a\neq b$ and by $(-1)^{\v i}
\phi^{ab}(\v i)$ for $a=b$.

\subsubsection{The classical helicity $\pm 2$ modes}

The two modes that have  a linear dispersion relation are the two
helicity $\pm 2$ modes. The dispersion relation is given by $\om_{\v
k} \sim \sqrt{\t g\t J} |\v k|$.  We note that the $\t U_2$ and $\t
U_1$ terms decouple from the helicity $\pm 2$ modes, since the
helicity $\pm 2$ modes satisfy the constraints
$\prt_i\phi^{ij}=R^{ii}=0$. Thus the dynamics of the helicity $\pm 2$
modes does not depend on $\t U_2$ and $\t U_1$.

To see if the result $\om_{\v k} \sim \sqrt{\t g\t J} |\v k|$ can be
trusted, we need to analyze the quantum fluctuations of $a_{ij}$ and
$\phi^{ij}$.  We find that for large $n_G$, the quantum fluctuations
of the helicity $\pm 2$ modes is of order $\del \phi^{ij},\del
a_{ij}\sim \sqrt{1/n_G}$ (note that $\t J$ and $\t g$ are of the same
order).  So the fluctuations of $\phi^{ij}$ and $a_{ij}$ satisfy
$1/n_G\ll \del \phi^{ij}\ll 1$, $1/n_G\ll \del a_{ij}\ll 1$ and the
semiclassical approximation is valid for the helicity $\pm 2$ modes.
In this case, the result $\om_{\v k}\sim \sqrt{\t g\t J}|\v k|$ can be
trusted.  The spin model \eq{HUJg} has gapless gravitons as its low
energy excitations.

\subsubsection{The quantum freeze and the gapping of helicity $0$ and $\pm 1$
modes}

The helicity $\pm 1$ modes and one of the helicity $0$ mode are
described by $(\th_i,\phi_i)$ which correspond to fluctuations  of
the following form: $a_{ij}= \prt_i \th_j+\prt_j \th_i$ and
$\phi^i=\prt_j \phi^{ji}$.  Their frequency $\om_{\v k}=0$.  For
such modes, the Hamiltonian only contains $\phi^i$. Thus the quantum
fluctuations satisfies $\del \phi^{i} \ll 1/n_G$ and $\del
\th_{i}\gg 1$. So the semiclassical approximation is not valid and
the result $\om_{\v k}=0$ cannot be trusted.  Using the similar
argument used in the emergence of $U(1)$ gauge bosons in section
\ref{clssU1}, we conclude that those modes are quantum frozen and
are gapped.  Since discreteness of $\phi^{ij}$ is of order $1/n_G$,
the gap of the helicity $0$ and $\pm 1$ modes is of order $\t
U_1/n_G$.  The strong fluctuations $\del a_{ij}= \prt_i \th_j+\prt_j
\th_i \gg 1$ lead to gauge transformations
\begin{equation}
\label{G1N}
 a_{ij}\to a_{ij}+ \prt_i \th_j+\prt_j \th_i
\end{equation}
and the weak fluctuations $\phi^i\ll 1/n_G$ lead to constraints
\begin{equation}
\label{C1N}
 \prt_j \phi^{ji}=0
\end{equation}

The second  helicity $0$ mode is described by $(\th,\phi)$ which
correspond to the following fluctuations: $\phi^{ij}=
(\del_{ij}\prt^2-\prt_i\prt_j) \phi$ and $\th=
(\del_{ij}\prt^2-\prt_i\prt_j)a_{ij}$.  Its frequency is again
$\om_{\v k}=0$. The Hamiltonian for such a mode contains only $\th$.
So the quantum fluctuations satisfies $ \del \phi \gg 1$ and $\del
\th\ll 1/n_G$. The second helicity $0$ mode is also gapped with a gap
of order $\t U_2$.  The strong fluctuations $\del \phi^{ij}=
(\del_{ij}\prt^2-\prt_i\prt_j) \phi \gg 1$ lead to a gauge
transformation
\begin{equation}
\label{G2N}
\phi^{ij}\to \phi^{ij}+(\del_{ij}\prt^2-\prt_i\prt_j) \phi
\end{equation}
and the weak fluctuations $ \th= (\del_{ij}\prt^2-\prt_i\prt_j)a_{ij}
\ll 1/n_G$ lead to a constraint
\begin{equation}
\label{C2N}
(\del_{ij}\prt^2-\prt_i\prt_j)a_{ij}=0
\end{equation}

The Lagrangian \eq{LagthphiN} equipped with the gauge transformations
(\ref{G1N},\ref{G2N}) and the constraints (\ref{C1N},\ref{C2N}) is
nothing but the linearized Einstein Lagrangian of gravity, where
$a_{ij}\sim g_{ij}-\del_{ij}$ represents the fluctuations of the
metric tenor $g_{ij}$ around the flat space. So the linearized Einstein
gravity may emerge from the quantum model \eq{HUJg} in the large
$n_G$ limit.  The qubit model \eq{HUJg} may be
quantum theory of gravity at the linear order.

The key in our argument is the gapping of helicity $0$ and $\pm 1$ modes.
Clearly, in the N-type lattice model \eq{HUJg}, those modes are strongly
fluctuating and strongly interacting modes. Those modes also have a very narrow
band width and arise from compact degree freedom on lattice.  The strongly
interacting modes with flat bands are in general gapped.  The gapless helicity
$\pm 2$ modes are very classical in the large $n_G$ limit. They should survive
the gapping of the  helicity $0$ and $\pm 1$ modes.

\subsection{Difficulties with the N-type lattice model}

\kp{\item
Since $[H_U,H_J]\neq 0$, and $[H_U,H_g]\neq 0$, the low
energy dynamics of the N-type model is hard to obtain
reliably.  The spin-wave/quantum-freeze
approach is an uncontrolled approximation, since the terms
beyond the quadratic order are not unimportant.
}

We have seen that the helicity $\pm 2$ modes are very classical with
small quantum fluctuations while the helicity $0$ and $\pm 1$ modes
have strong quantum fluctuations.  This is why the helicity $\pm 2$
modes remain gapless while the helicity $0$ and $\pm 1$ modes are
gapped.  Also, at quadratic level, the helicity $\pm 2$ modes and the
helicity $0,\pm 1$ modes decouple.  The strong quantum fluctuations in
the $0,\pm 1$ modes will not affect the dynamics of the $\pm 2$ modes
at the  quadratic level.

However, due to the strong quantum fluctuations in the  helicity
$0,\pm 1$ modes, the quadratic expansion for those modes is no
longer valid.  Beyond the quadratic level, the helicity $\pm 2$
modes and the helicity $0,\pm 1$ modes are coupled in the N-type
lattice model.  In this case, the strong quantum fluctuations in the
$0,\pm 1$ modes may change the dynamics of the $\pm 2$ modes beyond the
quadratic level. This is why we cannot obtain the low energy
dynamics of the N-type lattice model reliably and the $\om\sim |\v
k|$ dispersion for the helicity $\pm 2$ modes in the N-type lattice
model is not a reliable result.

However, such problem does not appear in the L-type lattice model.
The helicity $\pm 2$ modes and the helicity $0,\pm 1$ modes decouple
even beyond the quadratic level, since $H_U$ commutes with
$H^L_J+H^L_g$ in the L-type lattice model \eq{HUJgL}.  So the
$\om\sim |\v k|^3$ dispersion for the helicity $\pm 2$ modes in the
L-type lattice model is a reliable result.

\section{Topological defects}

In section \ref{lowspace}, we have constructed the low energy Hilbert
space of our lattice model in the large $\t U_{1,2}$ limit. We can
also construct high energy states that do not satisfy the scaler or
vector constraints (\ref{constr1}) but remain to be eigenstates of
$S(\v i)$ and $Q(\v i, \v i+\v a)$.  These states can be naturally
regarded as scalar defect and vector defect as discussed in the
continuum limit in section \ref{tensor}.

For example:
\begin{eqnarray}
|m, {\v i}\rangle=\left(\sum_{n=0}^{n_G} \exp\left[i2\pi
n\frac{m}{n_G}\right]S(\v i)^n\right) |\t 0\rangle
\end{eqnarray}
is a scalar defect with
\begin{eqnarray}
 S(\v j)|m, {\v i}\rangle&=&\delta_{\v i , \v
j}\exp\left[2\pi i\frac{m}{n_G}\right]|m, {\v i}\rangle\nonumber\\
m&=&0,1,2\cdots n_G-1
\end{eqnarray}

The vector defect can be constructed as, for example:
\begin{eqnarray}
|q_x, {\v i}\rangle &=& \left(W_\theta^{xx}(\v i)\right)^{q_x}|\t
0\rangle  \nonumber\\
|l_{xy}, {\v i}\rangle &=& \left(W_\theta^{xy} (\v i+\frac{\v
x}{2}+\frac{\v y}{2} \right)^{l_{xy}}|\t 0\rangle\nonumber\\
q_x,l_{xy}&=&0,1,2\cdots n_G-1
\end{eqnarray}
with
\begin{eqnarray}
&&Q(\v j, \v j+\v a)|q_x, {\v i}\rangle \nonumber\\
&=&\left(\delta_{\v i, \v j}\delta_{\v x, \v a}+\delta_{\v i-\v a,
\v j}\delta_{\v x, \v a}\right)\exp\left[2\pi i
\frac{q_x}{n_G}\right]|q_x, {\v i}\rangle  \\&&Q(\v j, \v j+\v
a)|l_{xy}, {\v i}\rangle \nonumber\\ &=& \left(\delta_{\v i, \v
j}\delta_{\v x, \v a}+\delta_{\v i, \v j}\delta_{\v y, \v
a}+\delta_{\v i+\v y, \v j}\delta_{\v x, \v a}+\delta_{\v i+\v x ,\v
j}\delta_{\v y, \v a}\right)\times \nonumber\\&& \exp\left[2\pi i
\frac{l_{xy}}{n_G}\right]|l_{xy}, {\v i}\rangle
\end{eqnarray}
notice the vector defect can only be created or annihilated in pairs.
Thus the vector defect in our lattice model has a
fractionalized nature.

Here the operator $\left(\sum_{n=0}^{n_G} \exp\left[i2\pi
n\frac{m}{n_G}\right]S(\v i)^n\right)$ and $\left(W_\theta^{xx}(\v
i)\right)^{q_x},\left(W_\theta^{xy} (\v i+\frac{\v x}{2}+\frac{\v
y}{2} \right)^{l_{xy}}$ can be viewed as scalar and vector defects
creating operators. The quantum number $m, q_{a}, l_{ab}$ divide the
Hilbert space into different sectors, our low energy Hilbert space
is defined as $m=q_{a}=l_{ab}=0$.

The existence of (quantized) scalar and vector defects have very
deep relationship with the existence of helicity $\pm 2$ modes in
low energy Hilbert space. In the continuum limit, it is easy to see
these defects are all topological defects and could not be removed
by local operations and protected by energy scale $\t U_1$ and $\t
U_2$. Meanwhile, this energy scales "freeze" the low energy quantum
fluctuations of helicity $0$ and $\pm 1$ modes. At this stage, we
may conclude the helicity $\pm 2$ modes in our lattice model is
topologically stable and the existence of scalar and vector defects
reflects the special quantum order in such a system.

\section{Conclusion and discussion}

In this paper, we studied two qubit models, the L-type model and the
N-type model.  The low energy dynamics of the L-type model can be
calculated reliably and we find that the helicity $\pm 2$ modes are
the only gapless excitations with $\om \sim |\v k|^3$ dispersion.  The
low energy dynamics of the N-type lattice model cannot be calculated
reliably. However, the spin-wave/quantum freeze approach suggest that
the helicity $\pm 2$ modes may be the only gapless excitations with
$\om \sim |\v k|$ dispersion in the N-type model.  Those gapless
helicity $\pm 2$ excitations can be viewed as emergent gravitons.

Both qubit models are constructed by putting a symmetric tensor field
on lattice, and by discretizing and compactifying each component of
the tensor field.  To understand the low energy dynamics of the two
model, we have used a semiclassical approach, where we treat the
tensor field as continuous and uncompactified.  We then expand the
lattice Lagrangian of the two models to quadratic order.  The
resulting free theories can be analyzed easily.  We find that both
models contain gapless helicity $\pm 2$ modes.  The analysis of the
quantum fluctuations for such modes indicates that the helicity $\pm
2$ modes are very classical and the semiclassical approach is
self consistent.  We also find that both models contain gapless
helicity $0,\ \pm 1$ modes.  The analysis of the quantum fluctuations
for those modes indicates that the helicity $0,\ \pm 1$ modes are
extremely quantum and the semiclassical approach is not
self consistent. The non-linear effects of the helicity $0,\ \pm 1$ modes
need to be considered.

However, the non-linear effects of the helicity $0,\ \pm 1$ modes have
very different forms in the L-type and the N-type model.  In the
L-type model, the strong fluctuating helicity $0,\ \pm 1$ modes
decouple from the semiclassical helicity $\pm 2$ modes even at
non-linear level. After taking into account the discreteness and the
compactness of the tensor field, we can show that non-linear effect
gaps the  helicity $0,\ \pm 1$ modes.  Such a gapping process does not
affect the helicity $\pm 2$ modes. So helicity $\pm 2$ modes remain
gapless with $\om \sim k^3$ dispersion.

In the N-type model, the strong fluctuating helicity $0,\ \pm 1$ modes
start to couple to the semiclassical helicity $\pm 2$ modes at
non-linear level.  After taking into account the discreteness and the
compactness of the tensor field, we can argue that non-linear
effect gaps the  helicity $0,\ \pm 1$ modes.  But such a gapping
process may affect the low energy dynamics of the helicity $\pm 2$
modes.  This is the reason why our result for the N-type model is not
reliable.

The situation for the emergence of the $\om\sim k$ gravity from the N-type
model is very similar to the emergence of Chern-Simons theory from
local qubit models in 1+2D.  Although we have no reliable analytic
calculation to show that the Chern-Simons theory can emerge from qubit
models, numerical calculations have convincingly established the
emergence of Chern-Simons theory from certain qubit models.  So
similarly, it is possible that the N-type model can produce the
linearized Einstein gravity, even though we can not show it reliably.
Numerical calculations are needed to show it convincingly.

We have seen that the gapping of the helicity $0$ mode in the rotor
model \eq{lattLC} leads to an emergence of $U(1)$ gauge structure at
low energies. The emergence of a gauge structure also represents a new
kind of order -- quantum order\cite{Wqoslpub,Wqoem} -- in the ground
state. In \Ref{HWcnt}, it was shown that the emergent $U(1)$ gauge
invariance, and hence the quantum order, is robust against any local
perturbations of the rotor model.  Thus the gaplessness of the
emergent photon is protected by the quantum order.\cite{Wlight}

Similarly, the gapping of the two helicity $0$ modes and the helicity
$\pm 1$ modes in the bosonic model (or qubit model) \eq{HUJgL} and/or
\eq{HUJg} leads to an emergent gauge invariance of the linearized
coordinate transformation. This indicates that the ground state of the
qubit models contain a new kind of quantum order (which is similar but
different from those associated with emergent ordinary gauge
invariance of internal degrees of freedom).  We expect such an
emergent linearized diffeomorphism invariance to be robust against any
local perturbations of the bosonic model. Thus the gaplessness of the
emergent helicity $\pm 2$ modes are protected by the quantum order and
are robust against any local perturbations.

Let us discuss the robustness of the $\om \sim  k^3$ dispersion of
the L-type model in more detail.  As stressed before, $\om \sim
k^3$ dispersion is the direct result of the local gauge invariance
under \eqn{T2} and \eqn{T11}.
To be more precise, we know that
the L-type model is described by the following effective Lagrangian
\begin{align}
\label{LLtype1}
 \cL&=\cE^{ij}\prt_0 a_{ij}
-\frac{J}{2} C^i_jC^i_j -\frac{g}{2} R^{ij}R^{ij} ...
\end{align}
Its Hamiltonian density is given by
$ \cH = \frac{J}{2} C^i_jC^i_j +\frac{g}{2} R^{ij}R^{ij} +...$,
where $...$ represent higher order and/or non-quadratic terms.  Such
an effective Hamiltonian density for the L-type model is directly
invariant under the gauge transformations \eq{T2} and \eq{T11}, which
leads to the $\om \sim  k^3$ dispersion.
Now, we would like to ask, if we integrate out
high energy fluctuations in the L-type model, can we generate
to following form of Hamiltonian density
\begin{align}
\del \cH &=
\frac{J'}{2}[{(\mathcal{E}^{ij})}^2 -\frac{1}{2}{(\mathcal{E}^{ii})}^2]
+\frac{g'}{2} a_{ij}R^{ij}
\nonumber\\
&=
\frac{J'}{2}[{(\mathcal{E}^{ij})}^2 -\frac{1}{2}{(\mathcal{E}^{ii})}^2]
+\frac{g'}{2} D^l_k D^k_l
\end{align}
that appears in the N-type model? Here $D^l_k = \eps^{ijl} \prt_i
a_{jk}$.  If such Hamiltonian density is generated, it will modify the
low energy dispersion from $\om \sim k^3$ to $\om \sim  k$.

Here we would like to stress that although $\del H=\int d^3\v x\, \del\cH$
is invariant under the gauge transformations \eq{T2} and \eq{T11}
(with the constraints \eq{T1} and \eq{T9}), the Hamiltonian density
$\del \cH$ is not invariant under the gauge transformations.  $\del
\cH$ changes by a total derivative.  Since $\cH$ is locally gauge
invariant, within a perturbative calculation, it cannot generate $\del
\cH$ that is only gauge invariant up to a surface term.  So the $\om
\sim  k^3$ dispersion of the L-type model is robust against all
perturbative corrections.  At the moment, we do not know any
non-perturbative effects that can generate the $\del \cH$ term within the
L-type model.

We further find the existence of scalar and vector defects to have
intrinsic relationship with the gapping of $0$ and $\pm 1$ modes. At
this stage, we may interpret the scalar defect as mass and vector
defects as momentum. They are all \emph{quantized} topological
defects. However, there's no direct evidence for mass quantization
experimentally, we attribute this fact to two possible reasons: a)
the mass quanta is too small to be detect experimentally; (of order
$\t U_2/n_G^2$ in the large $n_G$ limit) b) the appearance of $\om
\sim |\v k|$ gravitons may remove the mass quantization for low
lying excitations. To solve these problems demands much deeper
understanding of the origin of the linear dispersion relations and
will be our future efforts.

This research is supported by the Foundational Questions Institute (FQXi) and
NSF Grant DMR-0706078.

\appendix

\section{Low energy collective modes for L-type lattice model
-- equation-of-motion approach}
\label{eomLtype}

In this section, we will use a different approach --
equation-of-motion approach to study the low energy modes of the
L-type spin model \eq{HUJgL}.  The idea behind the equation-of-motion
approach is to choose operator that act within the low energy
constrained subspace, and find the equation of motion for the averages
of those operators. The derived equation of motion will describe the
classical motion of the low energy collective modes, provided that the
operators are chosen properly.  For our case we will choose the gauge
invariant operator $\cR_{ab}$ and $T^a_b$ to construct the equation of
the motion.

In the large $n_G$ limit, we can express $\exp(2\pi i/n_G)$ as
$1+2\pi i/n_G$ and the commutating relation becomes
\begin{align}
\label{com}
 W_L^{ab}W_\th^{ab}- W_\th^{ab}W_L^{ab}
\simeq\frac{2\pi i}{n_G}W_\th^{ab}W_L^{ab}
\end{align}
notice $T_a^b$ and $\cR_{ab}$ commute with the $H_U$ term, so the
equation of motions can be simplified as, for example:
\begin{widetext}

\begin{align}
&\partial_t \langle \cR_{yz}(\v i+\frac{\v y}{2}+\frac{\v z}{2})
\rangle = -i\langle[H,\cR_{yz}(\v i+\frac{\v y}{2}+\frac{\v
z}{2})]\rangle= -i\langle[H_J^L,\cR_{yz}(\v i+\frac{\v
y}{2}+\frac{\v z}{2})]\rangle=\pi J^L \times \nonumber\\& \langle
-T_x^x(\v i+\v y)-T_x^x(\v i-\v y)+T_x^x(\v i+\v z)+T_x^x(\v i-\v z)
-T_x^x(\v i-\v x+\v y)-T_x^x(\v i-\v x-\v y)+ T_x^x(\v i-\v x+\v z)
\nonumber\\& +T_x^x(\v i-\v x-\v z) -5T_y^y(\v i)-5T_y^y(\v i-\v
x)-T_y^y(\v i+\v x)-T_y^y(\v i-2\v x)-T_y^y(\v i+\v z)-T_y^y(\v i-\v
z)-T_y^y(\v i-\v x+\v z) \nonumber\\& -T_y^y(\v i-\v x-\v z)
+5T_z^z(\v i)+5T_z^z(\v i-\v x)+T_z^z(\v i+\v x)+T_z^z(\v i-2\v
x)+T_z^z(\v i+\v y)+T_z^z(\v i-\v y)+T_z^z(\v i-\v x+\v y)
\nonumber\\& +T_z^z(\v i-\v x-\v y) +\frac{1}{2}[-4T_x^y(\v
i)-4T_x^y(\v i+\v y)-T_x^y(\v i+\v z)-T_x^y(\v i-\v z)-T_x^y(\v i+\v
x)-T_x^y(\v i-\v x)-T_x^y(\v i+\v z+\v y) \nonumber\\& -T_x^y(\v
i-\v z+\v y)- T_x^y(\v i+\v x+\v y)-T_x^y(\v i-\v x+\v y) -4T_y^x(\v
i)-4T_y^x(\v i+\v y)-T_y^x(\v i+\v z)-T_y^x(\v i-\v z)-T_y^x(\v i+\v
x)\nonumber\\&-T_y^x(\v i-\v x) -T_y^x(\v i+\v z+\v y) -T_y^x(\v
i-\v z+\v y)-T_y^x(\v i+\v x+\v y)-T_y^x(\v i-\v x+\v y)  +4T_z^x(\v
i)+4T_z^x(\v i+\v z)+T_z^x(\v i+\v y)\nonumber\\&+T_z^x(\v i-\v y)+
T_z^x(\v i+\v x)+T_z^x(\v i-\v x)   +T_z^x(\v i+\v y+\v z)+T_z^x(\v
i-\v y+\v z)+ T_z^x(\v i+\v x+\v z)+T_z^x(\v i-\v x+\v z) +4T_x^z(\v
i) \nonumber\\&+4T_x^z(\v i+\v z)+T_x^z(\v i+\v y)+T_x^z(\v i-\v y)+
T_x^z(\v i+\v x)+T_x^z(\v i-\v x)  +T_x^z(\v i+\v y+\v z)+T_x^z(\v
i-\v y+\v z)+ T_x^z(\v i+\v x+\v z)\nonumber\\&+T_x^z(\v i-\v x+\v
z)]-h.c.\rangle \langle \cR_{yz}(\v i+\frac{\v y}{2}+\frac{\v z}{2})
\rangle
\end{align}

\begin{align}
&\partial_t \langle \cR_{zz}(\v i) \rangle= -i\langle[H,\cR_{zz}(\v
i)]\rangle = -i\langle[H_J^L,\cR_{zz}(\v i)]\rangle=-\frac{\pi
J^L}{2} \langle-2[T_x^x(\v i)+T_x^x(\v i-\v x)+T_x^x(\v i -\v
y)+T_x^x(\v i-\v z) \nonumber\\& +T_x^x(\v i-\v x-\v z)+T_x^x(\v
i-\v y-\v z)+T_x^x(\v i-\v x-\v y) +T_x^x(\v i-\v x-\v y-\v z)]
+2[T_y^y(\v i)+T_y^y(\v i-\v x)+T_y^y(\v i -\v y) \nonumber\\&
+T_y^y(\v i-\v z)+T_y^y(\v i-\v x-\v z)+T_y^y(\v i-\v y-\v
z)+T_y^y(\v i-\v x-\v y)+T_y^y(\v i-\v x-\v y-\v z)]
+\frac{1}{2}[T_x^y(\v i+\v y)+T_x^y(\v i-\v y) \nonumber\\&
-T_x^y(\v i+\v x) -T_x^y(\v i-\v x)+T_x^y(\v i-\v z+\v y)+T_x^y(\v
i-\v z-\v y)-T_x^y(\v i-\v z+\v x)-T_x^y(\v i-\v z-\v x) +T_y^x(\v
i+\v y) \nonumber\\& +T_y^x(\v i-\v y)-T_y^x(\v i+\v x)-T_y^x(\v
i-\v x)+T_y^x(\v i-\v z+\v y)+T_y^x(\v i-\v z-\v y)-T_y^x(\v i-\v
z+\v x)-T_y^x(\v i-\v z-\v x) \nonumber\\& +5T_y^z(\v i)+5T_y^z(\v
i-\v x)+T_y^z(\v i+\v x)+T_y^z(\v i-2\v x)+T_y^z(\v i+\v y)+T_y^z(\v
i-\v y)+T_y^z(\v i+\v y-\v x)+T_y^z(\v i-\v y-\v x) \nonumber\\&
+5T_z^y(\v i)+5T_z^y(\v i-\v x)+T_z^y(\v i+\v x)+T_z^y(\v i-2\v
x)+T_z^y(\v i+\v y)+T_z^y(\v i-\v y)+T_z^y(\v i+\v y-\v x)+T_z^y(\v
i-\v y-\v x) \nonumber\\& -5T_z^x(\v i)-5T_z^x(\v i-\v y)-T_z^x(\v
i+\v y)-T_z^x(\v i-2\v y)-T_z^x(\v i+\v x)-T_z^x(\v i-\v x)-T_z^x(\v
i+\v x-\v y)-T_z^x(\v i-\v x-\v y) \nonumber\\& -5T_x^z(\v
i)-5T_x^z(\v i-\v y)-T_x^z(\v i+\v y)-T_x^z(\v i-2\v y)-T_x^z(\v
i+\v x)-T_x^z(\v i-\v x)-T_x^z(\v i+\v x-\v y)-T_x^z(\v i-\v x-\v
y)] \nonumber\\& -h.c.\rangle\langle \cR_{zz}(\v i)\rangle
\end{align}

\begin{align}
&\partial_t \langle T_x^x(\v i)\rangle  = -i\langle[H,T_x^x(\v
i)]\rangle=-i\langle[H_g^L,T_x^x(\v i)]\rangle=\frac{\pi
g^L}{2}\langle T_x^x(\v i)\rangle\times \nonumber\\& \langle
2[\cR_{yy}(\v i)+\cR_{yy}(\v i+\v x)+\cR_{yy}(\v i+\v y)+\cR_{yy}(\v
i+\v z)\nonumber\\&+\cR_{yy}(\v i+\v x+\v y)+\cR_{yy}(\v i+\v y+\v
z)+\cR_{yy}(\v i+\v z+\v x)+\cR_{yy}(\v i+\v x+\v y+\v z)]
\nonumber\\& -2[\cR_{zz}(\v i)+\cR_{zz}(\v i+\v x)+\cR_{zz}(\v i+\v
y)+\cR_{zz}(\v i+\v z) \nonumber\\& +\cR_{zz}(\v i+\v x+\v
y)+\cR_{zz}(\v i+\v y+\v z)+\cR_{zz}(\v i+\v z+\v x)+\cR_{zz}(\v
i+\v x+\v y+\v z)] \nonumber\\& +5\cR_{xy}(\v i)+5\cR_{xy}(\v i+\v
z)+\cR_{xy}(\v i+2\v z)+\cR_{xy}(\v i-\v z)+\cR_{xy}(\v i+\v
y)+\cR_{xy}(\v i-\v y)+\cR_{xy}(\v i+\v z+\v y) \nonumber\\&
+\cR_{xy}(\v i+\v z-\v y) +\cR_{yz}(\v i+\v y)+\cR_{yz}(\v i-\v
y)-\cR_{yz}(\v i+\v z)-\cR_{yz}(\v i-\v z)+ \cR_{yz}(\v i+\v x+\v
y)+\cR_{yz}(\v i+\v x-\v y) \nonumber\\& -\cR_{yz}(\v i+\v x+\v z)
-\cR_{yz}(\v i+\v x-\v z) -5\cR_{zx}(\v i)-5\cR_{zx}(\v i+\v
y)-\cR_{zx}(\v i+2\v y)-\cR_{zx}(\v i-\v y)-\cR_{zx}(\v i+\v z)
\nonumber\\& -\cR_{zx}(\v i-\v z)-\cR_{zx}(\v i+\v y+\v
z)-\cR_{zx}(\v i+\v y-\v z)-h.c.\rangle
\end{align}

\begin{align}
&\partial_t  \langle T_y^x(\v i)\rangle  = -i\langle[H,T_y^x(\v
i)]\rangle=-i\langle[H_g^L,T_y^x(\v i)]\rangle=\pi g^L\langle
T_y^x(\v i)\rangle\times \nonumber\\& \langle -5\cR_{xx}(\v
i)-5\cR_{xx}(\v i+\v z)-\cR_{xx}(\v i+2\v z)-\cR_{xx}(\v i-\v
z)-\cR_{xx}(\v i+\v y)-\cR_{xx}(\v i-\v y)-\cR_{xx}(\v i+\v z+\v
y)-\cR_{xx}(\v i+\v z-\v y) \nonumber\\& +5\cR_{yy}(\v
i)+5\cR_{yy}(\v i+\v z)+\cR_{yy}(\v i+2\v z)+\cR_{yy}(\v i-\v
z)+\cR_{yy}(\v i+\v x)+\cR_{yy}(\v i-\v x)+\cR_{yy}(\v i+\v z+\v
x)+\cR_{yy}(\v i+\v z-\v x) \nonumber\\& +\cR_{zz}(\v i+\v
y)+\cR_{zz}(\v i-\v y)-\cR_{zz}(\v i+\v x) -\cR_{zz}(\v i-\v
x)+\cR_{zz}(\v i+\v z+\v y) \nonumber\\& +\cR_{zz}(\v i+\v z-\v
y)-\cR_{zz}(\v i+\v z+\v x)-\cR_{zz}(\v i+\v z-\v x) \nonumber\\&
+4\cR_{yz}(\v i)+4\cR_{yz}(\v i-\v y)+\cR_{yz}(\v i+\v
x)+\cR_{yz}(\v i-\v x)+\cR_{yz}(\v i-\v y+\v x)+\cR_{yz}(\v i-\v
y-\v x) \nonumber\\& +\cR_{yz}(\v i+\v z)+\cR_{yz}(\v i-\v
z)+\cR_{yz}(\v i-\v y+\v z)+\cR_{yz}(\v i-\v y-\v z) \nonumber\\&
-4\cR_{zx}(\v i)-4\cR_{zx}(\v i-\v x)-\cR_{zx}(\v i+\v
y)-\cR_{zx}(\v i-\v y)-\cR_{zx}(\v i-\v x+\v y)-\cR_{zx}(\v i-\v
x-\v y) \nonumber\\& -\cR_{zx}(\v i+\v z)-\cR_{zx}(\v i-\v
z)-\cR_{zx}(\v i-\v x+\v z)-\cR_{zx}(\v i-\v x-\v z)-h.c.\rangle
\end{align}

\begin{align}
&\partial_t \langle T_z^x(\v i)\rangle  = -i\langle[H,T_z^x(\v
i)]\rangle=-i\langle[H_g^L,T_z^x(\v i)]\rangle=\pi g^L\langle
T_z^x(\v i)\rangle \times \nonumber\\&\langle 5\cR_{xx}(\v
i)+5\cR_{xx}(\v i+\v y)+\cR_{xx}(\v i+2\v y)+\cR_{xx}(\v i-\v
y)+\cR_{xx}(\v i+\v z)+\cR_{xx}(\v i-\v z)+\cR_{xx}(\v i+\v y+\v
z)+\cR_{xx}(\v i+\v y-\v z) \nonumber\\& +\cR_{yy}(\v i+\v
x)+\cR_{yy}(\v i-\v x)-\cR_{yy}(\v i+\v z)-\cR_{yy}(\v i-\v z)
\nonumber\\& +\cR_{yy}(\v i+\v y+\v x)+\cR_{yy}(\v i+\v y-\v
x)-\cR_{yy}(\v i+\v y+\v z)-\cR_{yy}(\v i+\v y-\v z) \nonumber\\&
-5\cR_{zz}(\v i)-5\cR_{zz}(\v i+\v y)-\cR_{zz}(\v i+2\v
y)-\cR_{zz}(\v i-\v y)-\cR_{zz}(\v i+\v x)-\cR_{zz}(\v i-\v
x)-\cR_{zz}(\v i+\v y+\v x)-\cR_{zz}(\v i+\v y-\v x) \nonumber\\&
+4\cR_{xy}(\v i)+4\cR_{xy}(\v i-\v x)+\cR_{xy}(\v i+\v
y)+\cR_{xy}(\v i-\v y)+\cR_{xy}(\v i-\v x+\v y)+\cR_{xy}(\v i-\v
x-\v y) \nonumber\\& +\cR_{xy}(\v i+\v z)+\cR_{xy}(\v i-\v
z)+\cR_{xy}(\v i-\v x+\v z)+\cR_{xy}(\v i-\v x-\v z) \nonumber\\&
-4\cR_{yz}(\v i)-4\cR_{yz}(\v i-\v z)-\cR_{yz}(\v i+\v
y)-\cR_{yz}(\v i-\v y)-\cR_{yz}(\v i-\v z+\v y)-\cR_{yz}(\v i-\v
z-\v y) \nonumber\\& -\cR_{yz}(\v i+\v x)-\cR_{yz}(\v i-\v
x)-\cR_{yz}(\v i-\v z+\v x)-\cR_{yz}(\v i-\v z-\v x)-h.c.\rangle
\end{align}

\end{widetext}
other components of $\cR_{ab}$ and $T_a^b$ can be obtained by simply
cycling $xyz$ to $yzx$ and $zxy$.

In the large $n_G$ limit, the fluctuations $\Delta
\theta_{ab}\sim\Delta\phi^{ab}\sim 1/n_G$ is much less than $1$. So $\langle
T_a^b\rangle\sim\langle\cR_{ab}\rangle\sim1$(notice $ T_a^b$ and $\cR_{ab}$ are
unitary operators). One may express $\cR^{ab}$ as: $\langle\cR_{ab}\rangle=
e^{i\Theta_{ab}}$, where $\Theta_{ab}\sim 0$ is a small fluctuations around
$0$. This allow us use the approximation $\sin \Theta_{ab}\sim \Theta_{ab}$ and
linearized the above equation of motions in $\omega,k$ space.
Introducing
\begin{widetext}
\begin{eqnarray}
M=\left(\begin{array}{cccccc}0 & -2c_x c_y c_z & 2 c_x c_y c_z &
- c_z(c_y^2+c_z^2)&-c_x(c_y^2-c_z^2)&c_y(c_y^2+c_z^2) \nonumber\\
2c_x c_y c_z &0& -2 c_x c_y c_z &
 c_z(c_z^2+c_x^2)&- c_x(c_z^2+c_x^2)&
- c_y(c_z^2-c_x^2)
\nonumber\\
-2c_x c_y c_z & 2 c_x c_y c_z & 0 & - c_z(c_x^2-c_y^2)&
c_x(c_x^2+c_y^2)&
- c_y(c_x^2+c_y^2) \nonumber\\
2c_z(c_y^2+c_z^2)& -2c_z(c_z^2+c_x^2)& 2c_z(c_x^2-c_y^2)& 0
&-2c_y(c_z^2+c_x^2) & 2c_x(c_y^2+c_z^2) \nonumber\\
2c_x(c_y^2-c_z^2)& 2c_x(c_z^2+c_x^2)& -2c_x(c_x^2+c_y^2)
&2c_y(c_z^2+c_x^2) & 0 & -2c_z(c_x^2+c_y^2) \nonumber\\
-2c_y(c_y^2+c_z^2)& 2c_y(c_z^2-c_x^2)& 2c_y(c_x^2+c_y^2)
&-2c_x(c_y^2+c_z^2) & 2c_z(c_x^2+c_y^2) & 0
\end{array}\right)
\end{eqnarray}
\end{widetext}
where $c_x=2\cos\frac{k_x}{2}$,
$c_y=2\cos\frac{k_y}{2}$, $c_z=2\cos\frac{k_z}{2}$,
we can write the linearized equation of motions as
\begin{equation}
\omega^2\left(\begin{array}{c}\Theta_{xx}(\v{k})\nonumber\\
\Theta_{yy}(\v{k})
\nonumber\\ \Theta_{zz}(\v{k})\nonumber\\
\Theta_{xy}(\v{k})
\nonumber\\ \Theta_{yz}(\v{k})\nonumber\\
\Theta_{zx}(\v{k})
\end{array}\right)=-\pi^2 g^L J^L M^2 \left(\begin{array}{c}\Theta_{xx}(\v{k})\nonumber\\
\Theta_{yy}(\v{k})
\nonumber\\ \Theta_{zz}(\v{k})\nonumber\\
\Theta_{xy}(\v{k})
\nonumber\\ \Theta_{yz}(\v{k})\nonumber\\
\Theta_{zx}(\v{k})
\end{array}\right) \label{eqm}
\end{equation}
By solving the linearized
equation, we find that four of the six modes have zero frequency.
Those four modes have strong quantum fluctuations and will be gapped once we
take into account the compactness and discreteness of the fields $\Theta_{ab}$.
The only two low lying dynamical modes are helicity $\pm$ 2 modes with the
dispersion relations:
\begin{eqnarray}
\omega=2\pi \sqrt{8g^L
J^L\left(\cos^2\frac{k_x}{2}+\cos^2\frac{k_y}{2}+\cos^2\frac{k_z}{2}\right)^3}
\end{eqnarray}
This allows us to conclude that the low energy effective theory of
the lattice bosonic model \eq{HUJgL} is \eqn{LLtype} with the
constraints \eqn{T1} and \eqn{T9}. The low energy excitations are
helicity $\pm 2$ modes with $\om\propto k^3$ dispersion.


\end{document}